\documentclass[12pt]{article}

\usepackage{amsmath,amsfonts,amsbsy,amssymb,array,accents,dsfont}
\usepackage{enumerate,array,latexsym,graphicx,mathrsfs,verbatim,psfrag}
\usepackage{bm} 
\usepackage[normalem]{ulem}
\usepackage{booktabs} 
\usepackage[usenames]{color}
\usepackage[utf8]{inputenc}
\usepackage[english]{babel}
\usepackage{fancybox}

\usepackage{datetime}

\usepackage[nosort]{cite}
\usepackage{chngpage} 

\usepackage{enumitem}
\setlist{itemsep=0pt}

\usepackage[colorlinks=true,      linkcolor=darkblue,      urlcolor=darkblue,      
            filecolor=darkblue,      citecolor=darkblue,       pdfstartview=FitH,     
						pdfpagemode=UseNone,      bookmarksopen=true]{hyperref}  
\usepackage[all]{hypcap}     

\newcommand{\captionfonts}{\small}
\makeatletter  
\long\def\@makecaption#1#2{%
  \vskip\abovecaptionskip
  \sbox\@tempboxa{{\captionfonts #1: #2}}%
 \ifdim \wd\@tempboxa >\hsize
    {\captionfonts #1: #2\par}
  \else
    \hbox to\hsize{\hfil\box\@tempboxa\hfil}%
  \fi
  \vskip\belowcaptionskip}
\makeatother   

\DeclareMathSymbol{\medhatsym}{\mathord}{largesymbols}{"62} 

\DeclareMathSymbol{\medtildesym}{\mathord}{largesymbols}{"65}
\newcommand\lowermedtildesym{
  \text{\smash{\raisebox{-1.2ex}{%
    $\medtildesym$}}}}
\newcommand\medtilde[1]{
  \mathchoice
    {\accentset{\displaystyle\lowermedtildesym}{#1}}
    {\accentset{\textstyle\lowermedtildesym}{#1}}
    {\accentset{\scriptstyle\lowermedtildesym}{#1}}
    {\accentset{\scriptscriptstyle\lowermedtildesym}{#1}}
}

\numberwithin{equation}{section} 

\makeatletter
\g@addto@macro\bfseries{\boldmath}
\makeatother

\definecolor{cardinal}{rgb}{0.6,0,0}
\definecolor{darkgreen}{rgb}{0,0.4,0}
\definecolor{purple}{rgb}{0.5, 0, 0.5}
\definecolor{golden}{rgb}{0.92, 0.7, 0}
\definecolor{midnight}{rgb}{0, 0, 0.5}
\definecolor{darkblue}{rgb}{0, 0, 0.8}

\def\Neql#1{{\cal N}\!=\!{#1}}

\def\oneone{\rlap 1\mkern4mu{\rm l}}
\def\coeff#1#2{\relax{\textstyle {#1 \over #2}}\displaystyle}

\def\IC{\mathbb{C}}

\def\IR{\mathbb{R}}
\def\ZZ{\mathbb{Z}}

\def\cA{{\cal A}}
\def\cB{{\cal B}}
\def\cD{{\cal D}}

\def\cK{{\cal K}}
\def\cL{{\cal L}}
\def\cM{{\cal M}}

\def\cX{{\cal X}}
\def\cY{{\cal Y}}

\begin{document}
\begin{titlepage}


\bigskip
\bigskip
\bigskip
\bigskip
\centerline{\Large \bf  BPS equations and Non-trivial Compactifications}
\date{\today}
\bigskip
\bigskip
\centerline{{\bf Alexander Tyukov$^1$ and Nicholas P. Warner$^{1,2}$ }}
\bigskip
\centerline{$^1$ Department of Physics and Astronomy}
\centerline{and $^2$Department of Mathematics}
\centerline{University of Southern California} \centerline{Los
Angeles, CA 90089, USA}
\bigskip
\bigskip

\begin{abstract}
\noindent

\noindent 
We consider the problem of finding exact, eleven-dimensional, BPS supergravity solutions in which the compactification involves  a non-trivial Calabi-Yau manifold, $\cY$, as opposed to simply a $T^6$.   Since there are no explicitly-known metrics on non-trivial, compact Calabi-Yau manifolds, we use a non-compact ``local model'' and take the compactification manifold to be $\cY = \cM_{GH} \times T^2$ where $\cM_{GH}$ is a hyper-K\"ahler, Gibbons-Hawking ALE space.   We focus on backgrounds with three electric charges in  five dimensions and find  exact families of solutions to the BPS equations that have the same four supersymmetries as the three-charge black hole.  Our exact solution to the BPS system requires that the Calabi-Yau manifold be fibered over the space-time using compensators on $\cY$.   The role of the compensators is to ensure smoothness of the eleven-dimensional metric when the moduli of $\cY$ depend on the space-time.  The Maxwell field Ansatz also implicitly involves the compensators through the frames of the fibration.  We examine the equations of motion and discuss the brane distributions on generic internal manifolds that do not have enough symmetry to allow smearing.

\end{abstract}

\end{titlepage}

\tableofcontents

\section{Introduction}

Calabi-Yau manifolds, and other Ricci-flat compactifications, have played a major role in string theory since its resurgence in the 1980's.   If one is working with a low-energy, effective field theory in the uncompactified directions then the geometric details of such compactifying manifolds are often much less important than the topology.  Indeed, the cohomology structure determines not only the massless fields but also the Yukawa interactions.  On the other hand, the explicit metrics on almost all Calabi-Yau manifolds remain unknown, and the back-reaction of fields on the compactification manifold is typically ignored. 

At the other extreme, there are compactifications on tori and coset spaces, where detailed and explicit computations can be performed.  The problems here are that tori are often too trivial to lead to interesting physical models while coset spaces usually lead to AdS space-times, which are phenomenologically far less interesting than flat space.  Thus the Calabi-Yau manifold became the dominant method of creating physically interesting compactifications.

With the extensive developments in holographic field theory, the details of internal geometries once again became extremely important.  There are many examples of this, but perhaps the most celebrated is the conifold and its application to chiral symmetry breaking and confinement in holographic analogs of QCD \cite{Klebanov:2000hb}.  The original body of work, and the subsequent developments, have proven immensely influential and have shown how many of the geometric details of the conifold, and its deformations, play an essential role in describing the field-theory physics.   

In holographic applications, the ``compactifying'' manifold is not really compact:  at best, it is a cone over a compact manifold, with the radial coordinate along the cone providing the holographic renormalization scale.  In particular, the undeformed conifold is a cone over the coset manifold, $T^{(1,1)}$.  Such non-compact manifolds have also proven invaluable as ``local models'' of interesting regions of more traditional compactifications on Calabi-Yau manifolds.  (See \cite{Strominger:1995cz}, for an important early example.) Such ideas have also found extensive application in trying to create observationally-viable de Sitter cosmologies within string theory \cite{Kachru:2003aw,Kachru:2003sx}.  There has also been an animated debate about the viability of this mechanism \cite{Bena:2012bk,Blaback:2014tfa,Bena:2017uuz} and this analysis critically depends upon geometric details and the full back-reaction of fields on the conifold.

Our motivation for revisiting the problem of compactification on non-trivial Ricci-flat manifolds arises from the microstate geometry program.  The essential idea of this program is to replace black holes by smooth, horizonless geometries and then use these backgrounds to encode the microstate structure.  The back-reaction of all the fields and fluxes, sourced by  the detailed microstate structure, plays an essential role in this program because the goal is ensure that the complete solution is free of singularities and horizons.  One can then use holographic methods to analyze the encoding of  black-hole microstates.

Much of the early work on microstate geometries was done using the $T^6$ compactification of M-theory to five-dimensional supergravity coupled to two vector multiplets  \cite{Bena:2005va,Berglund:2005vb,Bena:2007kg}.  More recent work has been based on the $T^4$ compactification of IIB supergravity to six-dimensional $(0,1)$ supergravity coupled to two anti-self-dual tensor multiplets \cite{Bena:2011dd,Giusto:2013rxa,Bena:2015bea,Bena:2016agb, Bena:2016ypk,Bena:2017geu}.  The five-dimensional formulation was sufficient to find broad and interesting classes of microstate geometries and the move to the  six-dimensional formulation was driven by the need to describe much more detailed microstate structure.   In both instances, toroidal compactifications were sufficient to describe significant subsectors of the microstate structure using exact solutions of the supergravity equations of motion.  Moreover, the use of torus compactifications guarantees that the smooth solutions to the lower-dimensional supergravity uplift to smooth solutions of the higher-dimensional supergravity.

Recent work, using microstate geometries as backgrounds, suggests that a major fraction of black-hole microstate  structure can be encoded as condensates of W-branes \cite{Martinec:2015pfa,Martinec:2017ztd}.   One of the new facets  of W-branes is that  their behavior, and particularly their vast degeneracy, involves non-trivial sources and important dynamical details on the compactified directions.  Describing such microstate structure, with fully back-reacted detail,  requires going beyond the standard toroidal compactifications to more general Ricci-flat manifolds with non-trivial topology. 

More broadly,  there has been something of a disconnect between one geometric aspect of the microstate geometry program and the original, perturbative computations of black-hole degeneracies in string theory \cite{Strominger:1996sh,Maldacena:1997de}.   In the latter, most of the microstate structure and degeneracy is associated with non-trivial topology and its excitations within the compactification manifolds.  The early work on microstate geometries, particularly in five dimensions, used a trivial compactification manifold and all the interesting topology and dynamics lay in the space-time directions.  The more recent work on fluctuating microstate geometries, in both five and six dimensions, has begun  to bridge this divide in that the geometric fluctuations now make essential use of the compactification circles of the five-dimensional and six-dimensional geometries.   However, the tori that have been used to get these theories from M-theory and IIB supergravity still remain inert.  It seems natural to expect that the complete story of microstate geometries will treat the space-time topology and the compactification topology on an equal footing, and indeed weave both together non-trivially.   Moreover, the complete story should involve the dynamical fluctuations of all components of the fields both inside the space-time and inside the compactification manifold.  

This paper represents a very modest first step in this direction.  Our purpose is to see to what extent one can explicitly solve the BPS equations for very simple space-time geometries but with non-trivial, Ricci-flat compactification manifolds.  Ideally, the compactification manifold should be a Calabi-Yau manifold, or a K3 surface, but this immediately runs into the problem that the metrics are not explicitly known.   Instead we will use a ``local model:"  that is, we will replace the flat metric on $T^4$ by the Ricci-flat, hyper-K\"ahler metric on Gibbons-Hawking (GH) ALE space.  Put differently, we will replace a $T^4$ by $\IR^4$, make an orbifold of the $\IR^4$ and blow up that orbifold to create interesting local cohomology.   While not technically a compactification, we will continue to refer to the dimensional reduction of a theory on a non-compact, local model as a compactification.

Compactification on Gibbons-Hawking ALE spaces has been studied in other contexts, see, for example,  \cite{Minasian:2009rn}.  Here we  focus on the compactification of M-theory to five-dimensional, $\Neql2$ supergravity coupled to vector multiplets and we will replace the usual torus, $T^6$, by $\cM_{GH} \times T^2$, where $\cM_{GH}$ is a Riemannian\footnote{For microstate geometries with a GH metrics {\it in the space-time directions}, it is very important to allow this GH metric to be ambi-polar (see, for example, \cite{Bena:2007kg}) however, we will restrict our attention to Riemannian GH metrics in the internal, ``compactification'' directions.}  GH manifold.   Restricting to purely electrically charged objects in the space-time, we will analyze and solve the BPS equations for such a compactification.  The analysis is quite non-trivial because the compactifying manifold must be fibered over the space-time using compensator fields that play the role of gauge connections on the moduli space of $\cM_{GH}$.  Once one introduces the correct fiber structure, the solution of the BPS equations proceeds directly and leads to a  simple result.  

Solving the BPS equations does not always guarantee a solution to the equations of motion, particularly for backgrounds that only involve electric charge distributions.  Indeed, for the system we are considering, one must generically solve the  BPS equations, the Maxwell equations and the Bianchi identities in order to guarantee that  the Einstein equations are satisfied \cite{Gauntlett:2002fz}.  As we will discuss, in spite of the simplicity of the BPS system, the complete equations of motion are complicated because of the geometric details of $\cM_{GH}$. 

On the other hand, we will look at  the low-energy, effective, five-dimensional supergravity that emerges from a compactification on $\cM_{GH} \times T^2$ and investigate how the solutions to this system undergo uplifting to eleven-dimensional supergravity.   Such uplifts of five-dimensional solutions involve smearing branes over the internal manifold and precisely how this occurs, and the resulting brane distributions, will certainly depend upon the details of the non-trivial compactification geometry.   We will first look at this issue perturbatively in the GH moduli and then extend our analysis to finite values of the moduli.  For finite values of moduli, the translation symmetry of the compactification manifold is broken and so the branes do not smear out uniformly.  There are, in principle, choices of brane distributions on the internal space.  However, we examine the most canonical ``topological solution,'' for which the brane distribution on compactification manifold is always intrinsically dipolar  (with no net charge) outside $\delta$-function sources in the space-time.   Such dipolar charge distributions lead to fields that fall off extremely rapidly with distance and average to zero on scales that are much larger than the compactification scale.  The topological solution is, in this sense, the best one can do short of having the charge density vanish identically outside the $\delta$-function sources in the space-time. 

In Section  \ref{sect:Comp} we will first recall the details of the supersymmetry on the $T^6$ compactification of eleven-dimensional supergravity and then adapt this to a compactification on $\cM_{GH} \times T^2$.  In  Section  \ref{sect:MetAnsatz} we consider generic Calabi-Yau compactifications and how smoothness requires  one to fiber the Calabi-Yau manifold over the space-time using compensator fields as the fiber connections.  We begin  Section  \ref{sect:MGH} by computing the compensators for $\cM_{GH}$ and then we make, and solve, a BPS Ansatz for the metric and flux fields for eleven-dimensional supergravity compactified on $\cM_{GH} \times T^2$.  In Section  \ref{sect:EOMs} we discuss  BPS solutions to the equations of motion, including the linearized solution and  the ``topological solution.''  We look at the detailed example of an Eguchi-Hanson compactification in Section  \ref{sect:EH}.  Section  \ref{sect:Conc} contains our conclusions.

\leftline{\bf Notation and conventions:}   This paper will involve the reduction of eleven-dimensional supergravity on a six-dimensional, Riemannian internal manifold, $\cY$, to a five-dimensional  space-time, $\cX$.  In particular, $\cY$ will either be $T^6$ or $\cM_{GH} \times T^2$.  These spaces themselves will also involve spatial slices and fibrations, which will require further refinement of indices.  In particular, since we are dealing with BPS solutions, $\cX$ will come with a preferred time coordinate, $t$, and will find it convenient to write $\cX$ as time fibration over a spatial base, $\cB$.   To keep some order on this, we state many of our index conventions here.  First, the eleven-dimensional space-time and frame indices will be $M,N,P \dots $ and $A,B,C \dots$, respectively.  We will typically use the indices $\mu, \nu, \dots $ and $i, j, \dots$ for tangent indices on $\cB$ and $\cY$,  and these indices will range over $1,\dots,4$ and $1,\dots,6$,  respectively.   The  coordinates on $\cX$ and $\cY$ will be $(t, x^\mu)$ and $y^i$.   The  frame indices on $\cB$ will be  $\alpha, \beta, \dots$  taking values $1,\dots,4$ (with $0$ reserved for the time-like frame).  We will also typically use $a,b, \dots \in \{1,2,3,4\}$  to be frame indices  on $\cM_{GH}$. However, in Section \ref{sect:MetAnsatz}, we will consider compactifications on a generic manifold $\cX \times \cY$, where the detailed structure is not important. In this section, and only in this section,   $\mu, \nu, \dots $ and $\alpha, \beta, \dots$ will be tangent and frame indices on all of $\cX$ and take values  $0,1,\dots,4$.  Similarly, in Section \ref{sect:MetAnsatz}, and only in this section, we will take $a,b, \dots \in \{1,2, \dots, 6\}$ to be frame indices on a generic Calabi-Yau 3-fold, $\cY$.  

Finally, our M-theory and $\gamma$-matrix conventions are almost same as in \cite{Bena:2004de}, and we summarize them in Appendix \ref{app:conventions}

\section{ Compactifying M-theory to five dimensions}
\label{sect:Comp}

\subsection{The  standard ``STU'' compactification on $T^6$}
\label{ss:T6comp}

We start by recalling how the standard $\Neql2$ supergravity theory coupled to two vector multiplets in five dimensions (sometimes referred to as the five-dimensional ``STU'' model) can be  obtained from the $T^6$ compactification of  M-theory.    

The five-dimensional theory has three electromagnetic fields (one being the gravi-photon) and the electric charges correspond to three sets of M2 branes wrapping orthogonal $T^2$'s in the $T^6$.   The two scalars in the vector multiplets determine the relative scales of the three $T^2$'s.  The volume forms on the $T^2$ factors give rise to the five-dimensional vector fields via:
\begin{equation}
C^{(3)}  = A^{(1)} \wedge dy^1  \wedge dy^2 ~+~  A^{(2)}   \wedge
dy^3\wedge dy^4~+~ A^{(3)}  \wedge dy^5 \wedge dy^6 \,,
\label{C3field}
\end{equation}
where the vector fields, $A^{(I)}$, $I=1,2, 3$, are one-form Maxwell potentials on the five-dimensional space-time and depend only upon the coordinates, $x^\mu$.  We decompose these Maxwell fields into their electric and magnetic components:
\begin{equation}
A^{(I)}  ~=~  - Z_I^{-1} \, (dt +k) ~+~ C^{(I)}_{\hat \mu} \, d x^{\hat \mu}    \,.
\label{Aform1}
\end{equation}
Supersymmetry, or the BPS equations, require that the Maxwell electrostatic potentials are related, via the ``floating brane'' Ansatz \cite{Bena:2009fi}, to the warp factors appearing in the eleven-dimensional metric. Indeed, the M-theory metric has the form:
\begin{eqnarray}
 ds_{11}^2  = ds_5^2 & + &    \left(Z_2 Z_3  Z_1^{-2}  \right)^{1\over 3}
 ((dy^1)^2+(dy^2)^2) \nonumber \\
 & + & \left( Z_1 Z_3  Z_2^{-2} \right)^{1\over 3} ((dy^3)^2+(dy^4)^2)    +
  \left(Z_1 Z_2  Z_3^{-2} \right)^{1\over 3}  ((dy^5)^2+(dy^6)^2)) \,,
\label{elevenmetricT6}
\end{eqnarray}
with
\begin{equation}
ds_5^2 ~\equiv~ - \left( Z_1 Z_2  Z_3 \right)^{-{2\over 3}}  (dt+k)^2 +
\left( Z_1 Z_2 Z_3\right)^{1\over 3} \, ds_4^2 \,,
\label{fivemetric}
\end{equation}
for some spatial metric 
\begin{equation}
ds_4^2 ~\equiv~  h_{\mu \nu}dx^{\mu} dx^{\nu} \,.
\label{fourmetric}
\end{equation}

The supersymmetry, $\varepsilon$, of the solution  must satisfy the appropriate M2-brane projection conditions, which we take to be\footnote{The signs in the projectors  here are the same as those in \cite{Bena:2004de} but the opposite of those in \cite{Bena:2007kg}.  This is because there is an error in the latter reference. We are using projections that lead to the standard, five-dimensional BPS Ansatz and equations of \cite{Bena:2004de,Bena:2007kg} and that imply the projection condition (\ref{fourhelicity}), which leads to the {\it self-duality} constraint (\ref{halfflat}).}:
\begin{equation}
\big(\oneone ~+~  \Gamma^{056}) \, \varepsilon ~=~  \big(\oneone ~+~
\Gamma^{078}) \, \varepsilon ~=~  \big(\oneone ~+~  \Gamma^{09\,10}) \, \varepsilon ~=~  0  \,, 
\label{susyproj}
\end{equation}
where the frame indices $(5, \dots,10)$ correspond to the internal coordinates $(y^1, \dots, y^6)$.   These projection conditions define the supersymmetries of the three-charge, BMPV black hole.  Moreover, the supersymmetries of the corresponding microstate geometries obey precisely the same projection conditions (otherwise they could not represent microstates of the corresponding black hole).

Since the product of all the gamma-matrices is the identity matrix, as in (\ref{Gammaprod}), this implies
\begin{equation}
\big(\oneone ~-~  \Gamma^{1234} \big) \, \varepsilon ~=~   0  \,,
\label{fourhelicity}
\end{equation}
For the holonomy of the base metric, $ds_4^2$, on $\cB$ to preserve the supersymmetry, the metric must be hyper-K\"ahler with self-dual Riemann tensor: 
\begin{equation}
R^{(4)}_{\alpha \beta \gamma \delta }  ~=~ \coeff{1}{2}\, {\epsilon_{\gamma \delta}}^{\zeta \eta}\, R^{(4)}_{\alpha \beta \zeta \eta}    \,.
\label{halfflat}
\end{equation}
We also note that on the $T^2 \times T^2$ defined by $(y^1,y^2,y^3, y^4)$, (\ref{susyproj}) implies 
\begin{equation}
\big(\oneone ~+~ \Gamma^{5678}\big) \, \varepsilon ~=~  0  \,.
\label{T4helicity}
\end{equation}

If $\varepsilon$ is a supersymmetry then the vector defined by: 
\begin{equation}
\cK^M ~=~   \bar \varepsilon\,  \Gamma^M  \varepsilon  \,,
\label{Kvec1}
\end{equation}
is necessarily  a Killing vector  (see, for example, \cite{Gauntlett:2002fz}).  Moreover, suppose that $\varepsilon$ satisfies the projection condition:
\begin{equation}
\big(\oneone ~+~ \Gamma^{0AB}\big) \, \varepsilon ~=~  0  \,,
\label{genproj}
\end{equation}
for some $A, B$.  Then, by inserting $\Gamma^{0AB}$ into the right-hand side of (\ref{Kvec1}) and acting on both $\varepsilon$ and $\bar \varepsilon$, one can easily show that: 
\begin{equation}
\cK^C ~=~  \pm\, \cK^C   \qquad \Leftrightarrow \qquad  \Gamma^{C}\,  \Gamma^{0AB} ~=~  \pm \,\Gamma^{0AB}\, \Gamma^{C} \,.
\label{Kvecids}
\end{equation}
From  this, and the conditions (\ref{susyproj}), it follows that the only non-zero component of  $\cK^A$ is $\cK^0 =   \varepsilon^\dagger  \varepsilon$ and thus this Killing vector is necessarily time-like.

\subsection{Geometric preliminaries to reducing on $\cM_{GH} \times T^2$}
\label{ss:GHredn}

In the foregoing $T^6 = T^2 \times T^2\times T^2$ compactification, we are going to replace  $T^2 \times T^2$ factor by a more general manifold, $\medtilde M$, and the supersymmetry condition  (\ref{T4helicity}) means that the metric ${\medtilde {ds}}_4^2$ on  $\medtilde M$ must  be hyper-K\"ahler with an {\it anti-self-dual} Riemann tensor: 
\begin{equation}
\medtilde R^{(4)}_{abcd}  ~=~ -\coeff{1}{2}\, {\varepsilon_{cd}}^{ef}\, \medtilde R^{(4)}_{abef}    \,.
\label{halfflatint}
\end{equation}
Thus the metric on the internal manifold,  $\medtilde M \times T^2$, will  still be Ricci-flat and K\"ahler and so falls into the broader class of Calabi-Yau compactifications. We can therefore employ all the technology that has been developed for such compactifications.

The obvious choice for $\medtilde M$ is $K3$ but, since the metrics on such manifolds are not explicitly known,  we will take $\medtilde \cM$ to be a Gibbons-Hawking (ALE) space, $\cM_{GH}$, as a local model of a $K3$.  Since $\cM_{GH}$ is non-compact, it does not strictly represent a compactification, but we will abuse the terminology and still refer to it as a ``compactification.''  Despite the non-compactness of $\cM_{GH}$, we can still operate at the level of the equations of motion and use the BPS equations because these are all local and do not involve non-normalizable integrals over space. Indeed, at this level, one can equally replace $T^2 \times T^2$ by $\IR^2 \times \IR^2$.  When we consider effective actions, we will introduce a cut-off at large distances compared to the scales of the compact homology cycles of the GH manifold. 

We will therefore use the multi-centered, Riemannian GH metric with  {\it anti-self-dual} Riemann tensor:
\begin{equation}
	\medtilde{ds}_4^2 ~=~ V^{-1} (d\psi+A)^2 ~+~ V(d \vec z \cdot d \vec z)\,, 
	\label{GHmet}
\end{equation}
where $A =  \vec A \cdot d\vec z$ and
\begin{equation}
	\vec \nabla \times \vec A ~=~ - \vec \nabla V\,.
	\label{vectorA}
\end{equation}
We take the potential, $V$, to be positive definite\footnote{It would be extremely interesting to see if one can generalize this construction to ambi-polar GH metrics.} and define ``component functions,'' $K^I$, via:
\begin{equation}
	V ~=~ \sum_{I =1}^N \,K^I  \,, \qquad K^I ~\equiv~ \frac{q_I}{r_I}\,, \qquad r_I ~\equiv~ |\vec z - \vec z_I|\,,
	\label{VKIdefns}
\end{equation}
where $q_I  \in \ZZ_+$.   We also introduce the standard frames on the GH space:
\begin{equation}
\tilde{e}^1 ~=~ V^{-1/2}\, (d\psi+A) \,, \qquad \tilde{e}^{\hat a+1} ~=~ V^{1/2} dz^{\hat a}\,.
\label{GHframes}
\end{equation}
where we introduce indices $\hat a, \hat b, \dots  \in \{1,2,3\}$. With this choice of frames, and the choice (\ref{vectorA}), the spin connection and the curvature are anti-self-dual, as required.

The GH metric comes with three harmonic, self-dual K\"ahler forms: 
\begin{equation}
	J^{(\hat a)} ~=~ \Omega^{(\hat a)}_{+} ~\equiv~ \tilde{e}^1 \wedge \tilde{e}^{\hat a+1}  ~+~ \coeff{1}{2}\, \epsilon_{\hat a \hat b \hat c}\, \tilde{e}^{\hat b+1} \wedge \tilde{e}^{\hat c+1}\,.
\end{equation}

In a Calabi-Yau compactification of M-theory, the vector multiplets are associated with the harmonic $(1,1)$-forms of the compactification manifold.  Our goal is to get to five-dimensional, $\Neql2$ supergravity coupled to purely vector multiplets and so we have to choose a particular complex structure and then use it to identify the $(1,1)$-forms.   We choose the complex frames to be
\begin{equation}
	E^1~=~ \tilde{e}^1~+~ i\,\tilde{e}^4\,, \qquad E^2~=~ \tilde{e}^2 ~+~ i\,\tilde{e}^3 \,,
	\label{cplxframes}
\end{equation}
which means that  the K\"ahler form is given by
\begin{equation}
J ~=~ J^{(3)}  ~=~  \frac{i}{2 }(E^1\wedge \overline{E}^1 + E^2\wedge \overline{E}^2)  ~=~  \tilde{e}^1 \wedge \tilde{e}^4 + \tilde{e}^2 \wedge \tilde{e}^3  ~=~  (d\psi+A)\wedge dz^3 ~+~  V\, dz^1 \wedge dz^2 \,.
\label{Kform}
\end{equation}
The complex combinations,  $J^{(1)} + i\,J^{(2)}$ and  $J^{(1)} - i\,J^{(2)}$, are then harmonic $(2,0)$ and $(0,2)$ forms, respectively.  

In addition to complex structures, it is also useful to define three anti-self-dual two-forms 
\begin{equation}
	\Omega^{(\hat a)}_{-} ~=~ \tilde{e}^1 \wedge \tilde{e}^{\hat a+1} ~-~ \coeff{1}{2}\, \epsilon_{\hat a \hat b \hat c} \, \tilde{e}^{\hat b+1} \wedge \tilde{e}^{\hat c+1}\,.
\end{equation}
The harmonic forms on Gibbons-Hawking space with $N$ centers are given by: 
\begin{equation}
	\omega_{I} ~=~ \partial_{\hat a} \left(\frac{K^I}{V}\right) \Omega^{(\hat a)}_{-} = d \left[A^I - \frac{K^I}{V}(d\psi+A) \right]\,, 
	\label{harmforms}
\end{equation}
where
\begin{equation}
	\vec \nabla \times \vec A^I ~=~ - \vec \nabla K^I\,.
	\label{vectorAI}
\end{equation}
Note that these forms are anti-self dual, normalizable and dual to the homology cycles of the GH manifold. (See, for example, \cite{Bena:2007kg}\footnote{One should note that in this reference, the GH manifold is the base manifold of the space-time and the self-duality and anti-self-duality of the forms and curvatures are interchanged.}.)   Also observe that, because of  the first equation in (\ref{VKIdefns}), the $\omega_I$ satisfy the constraint:
\begin{equation}
	\sum_{I=1}^N \, \omega_I  ~=~0\,.
	\label{omconstr}
\end{equation}
There are thus only $(N-1)$ linearly independent such forms.

Since $J$ is self-dual and the $\Omega^{(\hat a)}_{-}$ are anti-self-dual, it follows from the structure of $SO(4)$ that the matrix ${J_a}^b$ commutes with the matrices ${\Omega^{(\hat a)}_{-}}{{}_a}^b$, where $a,b,... $ are frame indices on the GH space.  Hence:
\begin{equation}
{J_a}^c \, \Omega^{(\hat a)}_{-} {{}_c}^b ~=~ \Omega^{(\hat a)}_{-} {{}_a}^c  {J_c}^b  \qquad \Leftrightarrow \qquad   {J_a}^c \, {J_b}^d \, \Omega^{(\hat a)}_{-} {{}_{cd}}~=~ \Omega^{(\hat a)}_{-} {{}_{ab}}\,.
\label{11formproof}
\end{equation}
The second identity proves that all the harmonic forms given by (\ref{harmforms}) are, in fact, $(1,1)$-forms   with respect to $J$.  The complete set of $(1,1)$-forms on $\cM_{GH}$ is therefore spanned by linear combinations of  $\{J, \omega_I\}$ subject to the constraint (\ref{omconstr}). There are thus $N$ such harmonic forms, and therefore the reduction on $\cM_{GH}$ will produce $N$ massless vector fields, corresponding to $\Neql2$ supergravity coupled to $(N-1)$ vector multiplets. 

The analog of the $T^6$ projection conditions (\ref{susyproj}) can now be written on $\cM_{GH} \times T^2$ as\footnote{It turns that the second projector in (\ref{susyproj2}) is redundant as it is the square of the first one.  We include the second projector to simplify the exposition.}:
\begin{equation}
\Big(\oneone ~+~\frac{1}{4} \, J_{ab} \Gamma^{0\, a+4\, b+4} \,
\Big ) \, \varepsilon ~=~ \big(\oneone ~+~  \Gamma^{5678}) \, \varepsilon ~=~   \big(\oneone ~+~  \Gamma^{09\,10}) \, \varepsilon ~=~  0  \,.
\label{susyproj2}
\end{equation}

The metric, (\ref{GHmet}), has $3(N-1)$ moduli.  These are parametrized by the $\vec z_I$ in (\ref{VKIdefns}) with the overall translation of the center of mass of the points $\vec z_I$ being trivial. In principle, in a reduction on $\cM_{GH}$, all of these moduli will correspond to massless scalars in the space-time.  However, we are seeking a truncation with  $(N-1)$ vector multiplets, and so we need to specialize to $(N-1)$ of these moduli.  As we will discuss in Section \ref{sect:MGH}, the $(N-1)$ moduli that we seek are precisely those that preserve the complex structure defined in (\ref{Kform}) and they simply correspond to moving the GH points, $\vec z_I$, in the $z^3$-direction.

Finally, we recall that  the simplest, non-trivial GH space has $N=2$ and is simply the Eguchi-Hanson space.  There are thus two $(1,1)$ forms,  $J$ and $\omega$.   Replacing  $T^2 \times T^2$ by the Eguchi-Hanson space means that we replace torus 2-forms, $dy^1  \wedge dy^2+dy^3  \wedge dy^4$ and $dy^1  \wedge dy^2 - dy^3  \wedge dy^4$, by $J$ and $\omega$ respectively.  Like their torus counterparts, $J$ and $\omega$ also have trivial intersection: $J \wedge \omega = 0$.  Thus  compactification on the product of an Eguchi-Hanson space with a $T^2$ should lead to the standard ``STU'' model, and be indistinguishable from the $T^6$ compactification, in the low-energy, five-dimensional limit.

We now investigate how reduction on Calabi-Yau and GH manifolds works in more detail.

\section{Generalities about Calabi-Yau compactification}
\label{sect:MetAnsatz}

Our purpose here is to determine how the eleven-dimensional metric encodes the degrees of freedom of five-dimensional, $\Neql2$ supergravity coupled to $(N-1)$ vector multiplets.  In particular, the eleven-dimensional metric must encode the scalars of the vector multiplets as moduli of internal, ``compactification'' metric.  We start by considering the problem in a little more generality and take the internal manifold, $\cY$, to be Calabi-Yau, with coordinates $y^i$, Ricci-flat metric, $\medtilde{g}_{ij}$, and moduli, $u^I$.  We also take $\cX$ to be a generic space-time with coordinates, $x^\mu$.  Note that in this section, and only in this section,  $\mu, \nu, \dots $ and $\alpha, \beta, \dots$ will be tangent and frame indices on all of $\cX$ and take values  $0,1,\dots,4$ and that  $a,b, \dots \in \{1,2, \dots, 6\}$ will be frame indices of the entire manifold, $\cY$.   Also note that, in order to encode the vector multiplet scalars, the  metric on $\cY$ will be allowed to depend on $\cX$ but only through the moduli: $u^I = u^I(x^\mu)$.

\subsection{The form of the eleven-dimensional metric}
\label{ss:11met}

The naive choice for the eleven-dimensional metric is
\begin{equation}
	ds_{11}^2~=~ g_{\mu\nu}(x) dx^{\mu}dx^{\nu}+\medtilde{g}_{ij}(y,u(x))\,dy^{i}  \,dy^{j}   \,,
\label{naivemet}
\end{equation}
While (\ref{elevenmetricT6}) has this form, this is too simple an Ansatz for more general internal metrics.  As noted in \cite{Douglas:2008jx}, promoting the moduli to scalar fields on the space-time means that the metric must generically be fibered over the the space-time base, $\cX$.  We therefore take  the eleven-dimensional metric to have the form:
\begin{equation}
    ds_{11}^2~=~ g_{\mu\nu}(x) dx^{\mu}dx^{\nu}+\medtilde{g}_{ij}(y,u(x)) (dy^{i} - {B^{i}}_{\mu}(y,u(x)) dx^{\mu})(dy^{j} - {B^{j}}_{\nu}(y,u(x))dx^{\nu})\,.
\label{elevenmetfibered}
\end{equation}
The metric Ansatz in  \cite{Douglas:2008jx} omitted the quadratic terms in ${B^{i}}_{\mu}$ that appear in (\ref{elevenmetfibered}).  This was sufficient for the linearized action obtained in that  \cite{Douglas:2008jx}.  Our choice of (\ref{elevenmetfibered}) is motivated by the forms of fibrations that occur in consistent Kaluza-Klein Ans\"atze and, as we will see, our choice is essential to satisfying the eleven-dimensional BPS equations.

The fact that we are only going to allow the internal metric to depend on $x^\mu$ through the moduli also passes into the fibration in that the vector fields must have the form: 
\begin{equation}
    {B^{i}}_{\mu}(y,u(x))~=~   {B^{i}}_{I}(y,u(x)) \, \partial_{\mu} u^{I}(x)  \,.
\label{Bforms}
\end{equation}
Indeed the ${B^{i}}_{I}$ will be set equal to {\it compensators} on $\cY$, which are vector fields defined so to preserve regularity of the explicit metric components under derivatives with respect to the moduli.

\subsection{Compensators and Lichnerowicz modes}
\label{ss:Compensators}
 
Consider the family of smooth, Ricci-flat metrics on the internal manifold, $\cY$:
\begin{equation}
	\medtilde{ds}^2 ~=~ \tilde{g}_{i j} (y,u) \,  dy^i \, dy^j \,,
	\label{CYmet}
\end{equation}
which depends on some moduli, $u^I$.  By definition, changing $u^I$ moves (\ref{CYmet}) through a family of  smooth metrics, however  this does not mean that differentiating the coordinate dependent quantity, $\tilde{g}_{i j} (y,u)$, with respect to some $u^I$ will produce a smooth result.  What is guaranteed is that an infinitesimal shift in the $u^I$, {\it combined with an infinitesimal coordinate transformation,} will translate (\ref{CYmet}) horizontally across the space of gauge orbits of smooth metrics, moving from one smooth, Ricci-flat  metric to an infinitesimally neighboring smooth, Ricci-flat  metric. 

Thus we need make the combined transformation: 
\begin{equation}
u^I 	~\to~ u^I  ~+~ \delta u^I \,, \qquad  {y}^i ~\to~ y^i  + {B^{i}}_{I}(y,u) \, \delta u^I\,,
\end{equation}
for some appropriately-chosen {\it compensating vector} fields, ${B^{i}}_{I}(y,u)$.   Furthermore we define an associated covariant derivative 
\begin{equation}
	\cD_I ~=~ \frac{\partial}{\partial u^{I}}+\cL_{B_I}\,,
	\label{DIdefn}
\end{equation}
where $\cL_{B_I}$ is the Lie derivative along the compensating vector field ${B^{i}}_{I}$.  It is these fields that we will use in the metric Ansatz (\ref{elevenmetfibered}).

The fields ${B^{i}}_{I}(y,u)$ are chosen so as to ensure that 
\begin{equation}
 \delta_I   g_{i j} ~\equiv~ \cD_I  g_{i j} ~=~ \partial_I   g_{i j} + \cL_{B_I} g_{i j} ~=~ \partial_I  g_{i j} +  (\nabla_I B_{j\, I} + \nabla_j B_{i\, I})\,, 
 \label{LichHarm}
\end{equation}
is a smooth, symmetric tensor field.  This, however, does not fully specify ${B^{i}}_{I}$ and the canonical way to fix these vector fields is to require that the variations, $\delta_I   g_{i j}$, are transverse and traceless:
\begin{equation}
	\nabla^i \delta_I   g_{i j}~=~ 0\,, \qquad  g^{i j}\delta_I   g_{i j}~=~ 0\,. 
	\label{TTmodes}
\end{equation}
Since the family of metrics is Ricci-flat, the variations, $\delta_I   g_{i j}$, are also zero-modes of the Lichnerowicz operator, which, for transverse, traceless modes, can be written:
\begin{equation}
	\nabla_k \nabla^k \delta_I   g_{i j}~-~2  \, R_{k i j \ell}\, \delta_I   g^{k \ell} ~=~ 0\,. 
	\label{Lichnerowicz}
\end{equation}

On a four-dimensional K\"ahler manifold,  one can relate the harmonic $(1,1)$ forms, $\omega$, to  Lichnerowicz zero-modes obtained by variations of the K\"ahler moduli by:
\begin{equation}
	\delta  g_{i j}~=~ -\coeff{1}{2}\,\big( {J_i}^k \, \omega_{kj} ~+~ {J_j}^k \, \omega_{ki}  \big) \,. 
	\label{KahlerDefs}
\end{equation}

For hyper-K\"ahler manifolds one may use any of the three complex structures, which gives rise to three different moduli for each harmonic form.   For  GH manifolds, each of the $\omega_I$ in (\ref{harmforms}) corresponds, in this way, to the three-component position vectors, $\vec z_I$, in (\ref{VKIdefns}).  With the choice, $J=J^{(3)}$, in (\ref{Kform}), for the complex structure, the deformations of the form (\ref{KahlerDefs}), but involving $J=J^{(1)},J^{(2)}$, correspond to deformations of the complex structure.  As we will see in Section \ref{ss:GHcomps},  the choice of $J=J^{(3)}$, means that the third component, $z_I^3$ of $\vec z_I$ become the K\"ahler moduli while the other components, $z_I^{1,2}$ become complex structure moduli.

\subsection{Frames and spin connections}
\label{ss:frames}

We now return to our eleven-dimensional metric,  (\ref{elevenmetfibered}), and introduce frames and compute spin connections.  It is useful to start by introducing orthonormal frames for the individual metrics, $g_{\mu\nu}$ and $\medtilde{g}_{ij}$, in  (\ref{elevenmetfibered}):
\begin{equation}
\hat e^{\alpha} ~=~ {\hat e^{\alpha}}_{\mu}(x) \, dx^{\mu}   \,, \qquad \tilde{e}^{a}={\tilde{e}^{a}}_{i}(y,u)  \,dy^{i}  \,. 
	\label{frames1}
\end{equation}

We define ${\hat \omega}$ and $\tilde \omega$ to be the spin connections in these frames with the moduli, $u^I$, considered as free, constant parameters (not depending on coordinates):
\begin{equation}
  d\hat e^{\alpha}  ~=~ - {\hat \omega^{\alpha}}_{\beta}\wedge \hat e^{\beta}   \,, \qquad d \tilde e^{a}  ~=~ - {\tilde \omega^{a}}_{b}\wedge \hat e^{b}\,.
\label{ombits}
\end{equation}
We also define the individual components of these connections via:
\begin{equation}
{\hat \omega^{\alpha}}_{\beta}  ~=~  {{\hat \omega_\gamma}^{\alpha}}_{\beta} \, \hat e^{\gamma}   \,, \qquad {\tilde \omega^{a}}_{b}  ~=~  {{\tilde \omega_c}^{a}}_{b} \, \tilde e^{c} \,.
\label{ombitcomps}
\end{equation}
We will also find it convenient to introduce the restricted exterior derivatives:
\begin{equation}
d_{x}   ~\equiv~  dx^\mu \wedge \frac{\partial}{\partial x^\mu} \, \qquad d_{y}   ~\equiv~  dy^i \wedge \frac{\partial}{\partial y^i} \,.
\label{dXdY}
\end{equation}

We then take the frames of the fibered metric (\ref{elevenmetfibered}) to be:
\begin{equation}
 e^{\alpha} ~\equiv~ {e^{\alpha}}_{\mu}(x) dx^{\mu} ~=~ {\hat e^{\alpha}}_{\mu}(x) dx^{\mu}   \,, \qquad  e^{a}~=~ \tilde{e}^{a}- {B^{a}}_{\mu}dx^{\mu}  ~=~ {\tilde{e}^{a}}_{i}(y,u(x))dy^{i} - {B^{a}}_{\mu}dx^{\mu}\,. 
	\label{frames2}
\end{equation}
To write the spin connection explicitly, we define the tensors:
\begin{equation}
\begin{aligned}
{{F_I}^a}_b ~\equiv~ &  {\tilde{e}^{i}}_{b} \, \frac{ \partial {{\tilde e}^{a}}_{i}}{\partial u^I}  \,,  \qquad {{F_\mu}^a}_b ~\equiv~  {{F_I}^a}_b \,\partial_{\mu}u^{I}   \\
{{M_I}^a}_b ~\equiv~&  {\tilde{e}^{i}}_{b} \, \cD_I {{\tilde e}^{a}}_{i}~\equiv~{\tilde{e}^{i}}_{b} \, \frac{ \partial {{\tilde e}^{a}}_{i}}{\partial u^I} + \medtilde \nabla_b {B^a}_I \,,  \quad {{M_\alpha}^a}_b ~\equiv~ {\hat {e}^{\mu}}_{\alpha} \, \big( {{M_I}^a}_b \,\partial_{\mu}u^{I}  \big) \\
S_{\alpha \, ab}  ~\equiv~ & \coeff{1}{2} \big( M_{\alpha \, ab} + M_{\alpha\, ba} \big)  \,, \qquad A_{\alpha \, ab}   ~\equiv~  \coeff{1}{2} \big( M_{\alpha \, ab} - M_{\alpha\, ba} \big) \,, \\
 {Y^{i}}_{IJ}   ~\equiv~ & \frac{1}{2}\,   \bigg(  \frac{ \partial {B^{i}}_{J} }{\partial u_I} -   \frac{ \partial {B^{i}}_{I} }{\partial u_J}  + {B^{j}}_{I}  \, \partial_j  {B^{i}}_{J}  -   {B^{j}}_{J}  \,\partial_j  {B^{i}}_{I}\bigg) \,, \quad  {Y^{a}}_{\alpha\beta}  ~\equiv~ {\hat {e}^{\mu}}_{\alpha} \, {\hat {e}^{\nu}}_{\beta}\, {\tilde{e}^{a}}_{i} \,  {Y^{i}}_{IJ} \,  \partial_{\mu}u^{I}\,\partial_{\nu}u^{J}\,.   
\end{aligned}
\label{tensors1}
\end{equation}
One can also easily verify the following identity, which is useful in computing the spin connections:
\begin{equation}
 {Y^{a}}_{\alpha\beta}   ~\equiv~    {\hat {e}^{\mu}}_{[\alpha} \, {\hat {e}^{\nu}}_{\beta]} \,  \big(  \partial_\mu {B^{a}}_{\nu} +  {B^{b}}_{\mu}  \medtilde \nabla_b {B^{a}}_{\nu}  + {B^{b}}_{\mu} \, {{F_\nu}^a}_b \big) \,,
\label{tensors2}
\end{equation}
where $[\dots]$ represents skew-symmetrization of weight one.

A straightforward calculation leads to the following components of the spin connection:
\begin{align}
	{\omega^{\alpha}}_{\beta} &~=~ {{\omega_{\gamma}}^{\alpha}}_{\beta}\, e^{\gamma} +  {{Y_{a}}^\alpha}_\beta\, e^{a}\,,\nonumber \\
\label{spinconn}	{\omega^{a}}_{b} &~=~ {{{\tilde\omega}_{c}}{}^{a}}_{b} \, e^{c} - {{A_{\alpha}}^a}_b\, e^{\alpha}\,,   \\
		{\omega^{a}}_{\alpha} &~=~ {{S_{\alpha}}^a}_ b\, e^{b} - {Y^{a}}_{\alpha\beta} \, e^{\beta}\,. \nonumber
\end{align}

The symmetric part, $S_{I\, (a b)}$, of $M_{I\, (a b)}$  is, of course, the covariant derivative of the metric that leads to the Lichnerowicz mode:
\begin{equation}
	2\, S_{I\, (a b)}\, \tilde{e}^{a}_{i}\, \tilde{e}^{b}_{j} ~=~ \frac{\partial \medtilde{g}_{i j}}{\partial u^I} + (\medtilde \nabla_I B_{j\, I} + \medtilde \nabla_j B_{i\, I})  ~=~ \cD_I \medtilde  g_{i j}  ~=~  \delta_I  \medtilde g_{i j} \,. 
\label{lich2}
\end{equation}
 The anti-symmetric tensors, $A_{I\, a b}$, represent local Lorentz rotations of the frames that are induced by changes in the moduli.   The tensor ${Y^{i}}_{IJ}= -{Y^{i}}_{JI}$ may be thought of as  the field strength of the gauge fields ${B^i}_I$ {\it considered as tangent vectors on the moduli space.}  In this perspective, the vector index, $i$, tangent to $\cY$ is an internal  index and the last two terms in the definition of ${Y^{i}}_{IJ}$ in (\ref{tensors1}) make up the Lie derivative, $\cL_{B_I} B_J$, and so may be thought of as  structure constants in the algebra of the $B_J$.

The reason for the fibration structure in the metric Ansatz,  (\ref{elevenmetfibered}),  has begun to emerge from (\ref{tensors1})--(\ref{spinconn}).  Without the compensators, the spin connections would merely involve partial derivatives, with respect to the $u^I$, of the frames on $\cY$.  Such partial derivatives are generically singular.  For regularity, derivatives with respect to moduli must be paired with the corresponding compensator, as in (\ref{DIdefn}).  As is evident from the computation above, the compensators in the fibration structure achieves this pairing and results in a regular spin connection.  We will also see something similar with the flux Ansatz in Section \ref{ss:FluxAnsatz}.

\section{Reduction on  $\cM_{GH} \times T^2$ }
\label{sect:MGH}

Before we introduce our eleven-dimensional Ansatz and analyze the BPS equations for the compactification, we need a few more  geometric details of the internal GH manifold.  In particular, we need the compensators, ${B^{i}}_{I}$. Some of these results were obtained in  \cite{Schulz:2012wu} and we re-derive them  and a make a minor correction. We also revert to our previous convention in which $\mu, \nu \dots$ and  $\alpha, \beta, \dots$ are coordinate and  frame indices on $\cB$ (taking values $1,\dots,4$), $x^\mu$ are coordinates on $\cB$,  and $a,b, \dots $  are frame indices on $\cM_{GH}$, also taking values $1, \dots, 4$.

\subsection{Lichnerowicz modes, connections and compensators on $\cM_{GH}$}
\label{ss:GHcomps}

One can use (\ref{Kform}) and (\ref{harmforms}) in  (\ref{KahlerDefs}) to construct the  Lichnerowicz modes corresponding to K\"ahler deformations of the GH metric  (\ref{GHmet}).  One easily obtains:
\begin{equation}
	\begin{split}
	\delta_I \medtilde{ds}_4 ~\equiv~ 2\, S_{I\, ab} \, \tilde{e}^a \tilde{e}^b  ~=~ &\partial_3 \left(\frac{K^I}{V}\right) \left[ (\tilde{e}^1)^2+(\tilde{e}^4)^2-(\tilde{e}^2)^2-(\tilde{e}^3)^2\right]  \\  
  & ~+~  2\,  \partial_1 \left(\frac{K^I}{V}\right) \left[ \tilde{e}^2  \tilde{e}^4 - \tilde{e}^1 \tilde{e}^3 \right] + 
	2\,\partial_2 \left(\frac{K^I}{V}\right) \left[ \tilde{e}^1  \tilde{e}^2 + \tilde{e}^3  \tilde{e}^4 \right] \,,
	\end{split}
	\label{deform}
\end{equation}
where we have used this result to read off the tensors, $S_{I\, ab}$.  More explicitly, one has:  
\begin{equation}
	\omega_{I\, a b} ~=~ 2\, J_{a c} \, {S_{I}}{{}^{c}}_b  \,.
	\label{KahlerDefsGH}
\end{equation}

To obtain these modes  by differentiating the metric with respect to the moduli, $\vec z_I$, one must use the third components, $z_{I}^3$, combined with suitable compensators, ${B^{i}}_{I}$.  It is relatively easy to infer the form of the compensators by noting that the potential, $V$, defined in (\ref{VKIdefns}) develops a stronger singularity if one simply differentiates by $z_{I}^3$.  One can compensate for this by acting with an infinitesimal diffeomorphism that simultaneously moves the coordinate $z_3$ at  $\vec  z = \vec z_{I}$ but does not displace $\vec  z$ at  $ \vec z_{J}$ for $J\ne I$.  Such an infinitesimal diffeomorphism is generated by the vector field:
\begin{equation}
	B_I~=~  \frac{K^I}{V}\,\partial_3\,. 
\label{pieceB}
\end{equation}
Indeed, these compensator vector fields were given in  \cite{Schulz:2012wu} and they suffice if all the GH points lie along the $z^3$-axis.  However, when the GH points lie in general positions, one must allow the fiber coordinate to undergo and infinitesimal gauge transformation: $\psi \to \psi + f_I(\vec z)$ and thus the complete compensator has the form
\begin{equation}
	B_I~=~f_I (\vec z)\,\partial_{\psi} ~+~  \frac{K^I}{V}\,\partial_3\,. 
\label{CompForm}
\end{equation}
To fix the $f_I(\vec z)$ we simply compute $\cD_I \medtilde{ds}_4^2$, where $\cD_I$ is defined in (\ref{DIdefn}):
\begin{equation}
	\begin{split}
	\cD_I \medtilde{ds}_4^2  ~=~ & \partial_3 \left(\frac{K^I}{V}\right) \left[ (\tilde{e}^1)^2+(\tilde{e}^4)^2-(\tilde{e}^2)^2-(\tilde{e}^3)^2\right]  \\  
  & ~+~ 2\,  \partial_1 \left(\frac{K^I}{V}\right) \left[ \tilde{e}^2  \tilde{e}^4 - \tilde{e}^1  \tilde{e}^3 \right] + 
	2\,\partial_2 \left(\frac{K^I}{V}\right) \left[ \tilde{e}^1  \tilde{e}^2 + \tilde{e}^3  \tilde{e}^4 \right] \\
	&  ~+~ 2\, V^{-1/2} d_y \left(f_I + \frac{K^I}{V} A_3 - A^I_3 \right) \tilde{e}^1 \,.
	\end{split}
\end{equation}
Comparing this with  (\ref{deform}) we conclude that one has to choose $f_I = A^I_3 - \frac{K^I}{V} A_3$, where the $A^I$ are defined in (\ref{vectorAI}) and the complete compensators are therefore:
\begin{equation}
	B_I= \left(A^I_3 - \frac{K^I}{V} A_3 \right) \, \partial_{\psi} ~+~ \frac{K^I}{V} \, \partial_3\,.
	\label{compens}
\end{equation}
If  all GH centers lie on the $z^3$-axis then the third components, $A_3, A^I_3$, of the vector fields vanish identically and we recover the expression for compensators found in \cite{Schulz:2012wu}.

One can also act with $\cD_I$ on the K\"ahler form, $J$, and find the expected relation: 
\begin{equation}
	\cD_I  J ~=~ \omega_I\,. 
\end{equation}

We note, in passing, that since the K\"ahler deformations of the metric and complex structure are generated by $z_I^3$-derivatives, and not $z_I^{1,2}$-derivatives, this establishes that the $z_I^3$ are indeed the K\"ahler moduli and that the $z_I^{1,2}$ must be complex structure moduli.  Thus the $z_I^3$ represent scalars in the vector multiplets while the $z_I^{1,2}$ are scalars in  hypermultiplets.  We will allow the former to be dynamical while the latter will remain fixed.  The fact that we have obtained the compensators for generic locations of the GH points, rather than merely GH points along the $z^3$-axis, means that we can describe the dynamics in a background with generic vevs of the hypermultiplet scalars.

Finally,  the tensor, $F_{I\, a b}$,  defined in (\ref{tensors1}), has the following non-zero components:
\begin{equation}
	\begin{split}
	F_{I\, 1 1} ~=~ F_{I\, 4 4}  ~=~ - F_{I\, 2 2}  ~=~ - F_{I\, 3 3}  ~=~ \frac{1}{2} \, \partial_3 \left(\frac{K^I}{V}\right)\,,  \\
	F_{I\, 1 2}  ~=~ F_{I\, 4 3}  ~=~ \partial_2 \left(\frac{K^I}{V}\right)\,, \quad  
	F_{I\, 4 2}  ~=~ - F_{I\, 1 3}  ~=~ \partial_1 \left(\frac{K^I}{V}\right)\,.
	\end{split}
\label{Ftensor}
\end{equation}
Symmetrizing gives $S_{I\, (a b)}$ and the result matches with the expression (\ref{deform}).  Skew symmetrization leads to the components of the 
anti-symmetric tensor $A_{I \, a b}$:
\begin{equation}
	\frac12 A_{I\, a b} \, \tilde{e}^a \wedge \tilde{e}^b ~=~ \frac12 \partial_2 \left(\frac{K^I}{V}\right) \left[ \tilde{e}^1 \wedge \tilde{e}^2 - \tilde{e}^3 \wedge \tilde{e}^3 \right] - \frac12 \partial_1 \left(\frac{K^I}{V}\right) \left[ \tilde{e}^1 \wedge \tilde{e}^3 + \tilde{e}^2 \wedge \tilde{e}^4 \right] \,,
	\label{Ares}
\end{equation}
which is anti-self-dual.

It is not difficult to check that the tensor ${Y^i}_{I J}$ defined by (\ref{tensors1})  vanishes identically for the compensating fields (\ref{compens}). The  fields $B_I$ are, in this sense,  pure gauge.

\subsection{The metric Ansatz}
\label{ss:MetAnsatz}

We will take the full eleven-dimensional metric Ansatz to be a warped fibration of $\cM_{GH} \times T^2$ over a five-dimensional space-time, $\cX$.  We will require the metric on $\cX$ to be precisely those of the dimensionally reduced five-dimensional supergravity solutions.  On the internal manifold, there is a warp factor that determines the relative volume of the $\cM_{GH}$ and the $T^2$.  This scalar is part of a vector multiplet in five dimensions and so it must be allowed to depend non-trivially upon the coordinates, $x^\mu$, of $\cX$.  The other vector multiplet scalars are the $(N-1)$ K\"ahler moduli of $\cM_{GH}$ and so these can also depend on $x^\mu$.  The overall volume of   $\cM_{GH} \times T^2$ represents a scalar in the five-dimensional universal hypermultiplet and so this degree of freedom will be frozen. 

For simplicity, we will also assume that the background is purely electrically charged and has no angular momentum.  Our metric Ansatz is therefore: 
\begin{align}
	ds_{11}^2 &~=~ -Z^{-2}\, dt^2 + Z\, ds_4^2 +\left(\frac{Z}{Z_0}\right)\,\medtilde{ds}_4^2  + \left(\frac{Z_0}{Z}\right)^2(dy_5^2+dy_6^2)\,, \nonumber\\
	ds_4^2 &~=~ g_{\mu\nu}(x) \, dx^{\mu}dx^{\nu}\,, \label{11dmetric} \\
	\medtilde{ds}_4^2  &~=~ \medtilde{g}_{ij}\, (y,u(x)) (dy^{i} -{B^{i}}_{\mu}(y,u(x)) dx^{\mu})(dy^{j} - {B^{j}}_{\nu}(y,u(x))dx^{\nu})\,, \nonumber 
\end{align}
The functions $Z$ and $Z_0$ depend only on the  coordinates $x^\mu$ on $\cX$ and the BPS condition means that all the fields must, of course, be time-independent.  The factors of $Z$ in the space-time metric are motivated directly by (\ref{fivemetric}) and the fact that eleven-dimensional solution must reduce to that of five-dimensional supergravity.   There could, in principle, have been another warp-factor, $W(x^\mu, y^i)$, multiplying the overall, five-dimensional space-time metric.  However, we found that such a function was rendered trivial by the BPS conditions and so we have not included it here.  We have also considered other generalizations of this Ansatz but the ultimate justification of our choice (\ref{11dmetric}) is that we will show that it is sufficient to solve the BPS equations.

We introduce frames
\begin{equation}
e^0 = Z^{-1} \,dt\,,\quad e^{\alpha} = Z^{1/2} \, {\hat {e}^{\alpha}}_{ \mu} \, dx^{ \mu}\,,\quad  e^{a+4}= \left(\frac{Z}{Z_0}\right)^{\frac{1}{2}}{\tilde{e}^{a}}_{i}\,(dy^{i} - {B^{i}}_{\mu}\,dx^{\mu})\,, \quad
e^{9,10}= \left(\frac{Z_0}{Z}\right) dy^{5,6}\,,
\end{equation} 
for which the the spin connection is:
\begin{align}
	{\omega^{0}}_{\alpha} &~=~ -Z^{-1/2}\, (d \log Z)_{\alpha}\, e^0\,, \nonumber \\ 
	{\omega^{\alpha}}_{ \beta} &~=~  Z^{-1/2}\,{{\hat\omega_{\gamma}}^{\alpha}}_{\beta}\, e^{\gamma} +  Z^{-1/2}\,(d \log Z^{1/2})_{\beta} \,  e^{\alpha} - Z^{-1/2}\,(d \log Z^{1/2})_{\alpha} \, e^{\beta}  \nonumber \\ 
	& \qquad\qquad\qquad ~+~  Z^{-1/2}\, Z_0^{-1/2}\, Y_{a\,[\alpha \beta]}\, e^{a}\,,\nonumber \\
	{\omega^{a+4}}_{b+4} &~=~\left(\frac{Z_0}{Z}\right)\,{{\tilde{\omega}_{c}}{}^{a}}_{b} \, e^{c+4} - Z^{-1/2}\, A_{\alpha\,[a b]}\, e^{\alpha}\,, \\
		{\omega^{a+4}}_{\alpha} &~=~ Z^{-1/2}\, S_{\alpha\,a b}\, e^{b+4} + Z^{-1/2}\, \left(d\log (Z/Z_0)^{1/2}\right)_{\alpha} - Z^{-1/2}\, Z_0^{-1/2}\, {Y^{a}}_{\alpha \beta}\, e^{\beta}\,, \nonumber  \\
		{\omega^{9,10}}_{\alpha} &~=~ Z^{-1/2}\, \left(d\log (Z_0/Z)\right)_{\alpha}\, e^{9,10}\,, \nonumber
\end{align}
where $\hat\omega$ and $\tilde \omega$ are defined in  (\ref{ombits}) and (\ref{ombitcomps}).

\subsection{The flux Ansatz}
\label{ss:FluxAnsatz}

Since we are assuming that the background is purely electrically charged, the usual Calabi-Yau Ansatz suggests that we should take 
\begin{equation}
	C^{(3)} ~=~ - \Phi_0\, dt \wedge J ~+~ \sum_I \, \Phi_I \, dt \wedge \omega_I \,,
	\label{naiveC3}
\end{equation}
for some potential functions $\Phi_0, \Phi_I$.   This is too simplistic.  

First we need to use the compensators to ensure that $F^{(4)} = d C^{(3)}$ is suitably smooth and so,  as in the metric Ansatz, this means that we must fiber the K\"ahler form.    We therefore introduce the frame components of $J$ via:
\begin{equation}
J ~\equiv~ \coeff{1}{2} \, J_{ab} \, \tilde{e}^a \wedge \tilde{e}^b \,,
\label{Jcomps}
\end{equation}
and define the form $J'$ on $\cX \times \cY$ by:
\begin{equation}
	J' ~\equiv~  \coeff{1}{2} \, J_{a b}\, (\tilde{e}^a - {B^a}_{\alpha}\, \hat{e}^{\alpha}) \wedge (\tilde{e}^b -  {B^b}_{\beta}\, \hat{e}^{\beta}) = J + \coeff{1}{2} \, J_{a b}  {B^a}_{\alpha} {B^b}_{\beta} \, \hat{e}^{\alpha} \wedge \hat{e}^{\beta} - J_{a b} {B^b}_{\beta} \, \tilde{e}^a \wedge \hat{e}^{\beta} \,,
\end{equation}
where ${B^a}_{\alpha}  \equiv  {\tilde{e}^a}_i {\hat{e}^{\mu}}_{\alpha} {B^i}_I  \partial_{\mu} u^I$. One can also fiber the $\omega_I$ in a similar manner. 

The total exterior derivative of $J'$ can be computed using (\ref{spinconn})
 \begin{equation}
	 d\, J' = (d_x+d_y)\, J' =  J_{a b} \, {A_{\alpha}}{{}^{a}}_c \, \tilde{e}^c \wedge \tilde{e}^b \wedge \hat{e}^{\alpha} -  J_{a b} \,{S_{\alpha}}{{}^{a}}_c  \, \tilde{e}^b \wedge \tilde{e}^c \wedge \hat{e}^{\alpha} + J_{a b} \, {\tilde{\omega}_d}{{}^a}_{c} \tilde{e}^b \wedge \tilde{e}^c \wedge \tilde{e}^d \,. 	  
	 \label{dJprime}
 \end{equation}
Now recall that the spin-connection and ${A_{\alpha}}{{}^{a}}_c$ are anti-self-dual (see (\ref{Ares})) and that the  commutator of a self-dual matrix with an anti-self-dual one is zero.   It follows that the first and the third terms in  (\ref{dJprime}) vanish.  Indeed, the vanishing of the last term simply represents the closure of the K\"ahler form.  We therefore arrive at the simple result:
\begin{equation}
	d\, J' = J_{a c} \,{S_{\alpha}}{{}^{c}}_b \, \tilde{e}^a \wedge \tilde{e}^b \wedge \hat{e}^{\alpha} = \cD_I J \wedge du^I(x) 
	= \omega_I \wedge du^I(x)  \,.
\end{equation}
In particular, observe that the exterior derivative of $J'$ generates precisely the smooth harmonic forms wedged with the exterior derivatives of the vector multiplet scalars. 

This leads to a particularly simple Ansatz for the three-form potential 
\begin{equation}
\begin{aligned}
	C^{(3)} ~=~ &  - \coeff{1}{2}\, W^{-1}  \, J_{ab} \,dt \wedge e^{a+4} \wedge e^{b+4} ~-~ Z_3^{-1}\, dt \wedge dy^{5} \wedge dy^{6}    \\
	 ~=~  & -W^{-1} \, \left(\frac{Z}{Z_0}\right) \, dt \wedge J' ~-~ Z_3^{-1}\, dt \wedge dy^{5} \wedge dy^{6} \,,
\end{aligned}
	\label{C3Ansatz}
\end{equation}
for some potential functions, $W$ and $Z_3$.  The four-form flux is then:
\begin{align}
	F^{(4)}  ~=~& \coeff{1}{2}\,  W^{-1}  \, Z^{1/2} \, {e^\mu}_\alpha \,\left[ \omega_{I\, a b} \, \partial_{\mu} u^I -J_{a b} \, (\partial_\mu \log (W Z_0/Z))  \right] \, e^0 \wedge e^{\alpha} \wedge e^{a+4} \wedge e^{b+4} \nonumber \\
	& ~-~  Z_3^{-1} \,  Z^{1/2} \,  \left(\frac{Z}{Z_0}\right)^2 \, (\partial_\mu \log Z_3) \, e^0 \wedge e^{\alpha} \wedge e^{9} \wedge e^{10} \,,
	\label{F4flux}
\end{align}
where $\omega_{I\, a b} = 2 J_{a c} {S_{I}}{{}^{c}}_b $ are the coefficients of two-form $\omega_I$.    We have also considered adding explicit terms proportional to $\omega_I$ in the Ansatz (\ref{C3Ansatz}) for $C^{(3)}$, as in (\ref{naiveC3}).  However, solving the BPS equations eliminated these terms and showed that the simple Ansatz, (\ref{C3Ansatz}), and the $\omega_I$-terms that it generates in $F^{(4)}$, are sufficient.

\subsection{The BPS equations}
\label{ss:BPSeqns}

The BPS equations are given by the vanishing of the gravitino variations, which we write in frames:
\begin{equation}
	{e^M}_{A} \, \delta\psi_{M} ~=~ {e^M}_{A} \,\partial_M \varepsilon + \frac{1}{4}\,  \omega_{A B C} \Gamma^{B C} \varepsilon 
	+ \frac{1}{288}\, \left( {\Gamma_A}^{B C D E} F_{B C D E} - 8 \Gamma^{B C D} F_{A B C D} \right) \varepsilon ~=~ 0 \,.
	\label{gravvar}
\end{equation}
We will also impose the projection conditions (\ref{susyproj2}) on $\varepsilon$.  In particular, we note that the second projection in (\ref{susyproj2}) implies that 
for any anti-self-dual two-form, $\cA$, on internal four-fold one has: 
\begin{equation}
	\cA_{a b} \, \Gamma^{a+4\, b+4} \, \varepsilon ~=~ 0 \,.
	\label{asd}
\end{equation}

Assuming that Killing spinor, $\varepsilon$, is time independent, the $0$-component of BPS equations gives:
\begin{equation}
\begin{aligned}
		{e^M}_{0}\,\delta\psi_{M} ~=~ \frac{1}{2} \, Z^{-1/2}\,{e^\mu}_\alpha  \bigg[ (\partial_\mu \log Z)\, \Gamma^{0 \alpha}  &  ~-~  \frac{1}{6} \,\bigg( \omega_{I\, a b} \,\partial_{\mu} u^I - \frac{Z}{W} \,(\partial_\mu \log (W Z_0/Z)) \, J_{a b} \bigg)\,  \Gamma^{\alpha\, a+4\, b+4}  \\
&  ~+~  \frac{1}{3} \,\left(\frac{Z}{Z_0}\right)^2 \frac{Z}{Z_3} \,  (\partial_\mu \log Z_3)\, \Gamma^{\alpha 9 10} \bigg] \,  \varepsilon ~=~ 0 \,.
\end{aligned}
\end{equation}
Using the  projection conditions and  (\ref{asd}) to eliminate the $\omega_I$-term, reduces this to:
\begin{equation}
{e^\mu}_\alpha \, \big(\coeff{1}{2} \, Z^{-1/2}\,  \Gamma^{0 \alpha}  \,  \varepsilon\big)\, \bigg[\partial_\mu  \log Z ~-~  \frac{2}{3} \, \frac{Z}{W} \,\partial_\mu  \log (W Z_0/Z)
 ~-~  \frac{1}{3} \, \frac{Z^3}{Z_0^2 \, Z_3}  \,  \partial_\mu   \log Z_3\, \bigg]   ~=~ 0 \,.
 \label{vareqn1}
\end{equation}
Similarly, the component of (\ref{gravvar}) parallel to the spatial sections of $\cX$ gives the equation:
\begin{equation}
\begin{aligned}
 &Z^{-1/2}\, {e^\mu}_\gamma \big[  \partial_\mu +  \coeff{1}{6}\,  \partial_\mu  \log Z_3  +  \coeff{1}{3}\,  \partial_\mu  \log Z_0  \, \big] \,\varepsilon \\ 
&~+~    \coeff{1}{12}\, Z^{-1/2}\,   \big( {e^\mu}_\alpha \, {\Gamma_\gamma}^{\alpha} \,\varepsilon\big)\, \big[3 \,  \partial_\mu  \log Z - 2 \,  \partial_\mu  \log Z_0 -   \partial_\mu  \log Z_3\, \big]   ~=~ 0 \,,
 \label{vareqn2}
 \end{aligned}
\end{equation}
where we have used   the projection condition (\ref{fourhelicity}) combined with the self-duality of the spin-connection,  $\hat \omega_{\hat \alpha \hat \beta}$, on $\cX$ to eliminate the spin connection terms in  (\ref{vareqn2}).

The component parallel to $y^5$ or $y^6$ leads to 
\begin{equation}
- \coeff{1}{6}\, Z^{-1/2}\,   \big( {e^\mu}_\alpha \, {\Gamma}^{A\,\alpha} \,\varepsilon\big)\, \big[3 \,   \partial_\mu  \log Z ~-~  2\,    \partial_\mu  \log Z_0 ~-~   \partial_\mu  \log Z_3 \, \big]   ~=~ 0 \,,
 \label{vareqn3}
\end{equation}
for $A =9,10$.

As a result of (\ref{vareqn1}) -- (\ref{vareqn3}), we see that 
\begin{equation}
Z ~=~(Z_0^2 \, Z_3)^{1/3}\,, \qquad  W   ~=~ Z  \,,
 \label{resvar1}
\end{equation}
and
\begin{equation}
\varepsilon ~=~Z^{-1/2} \, \varepsilon_0  \,,
 \label{resvar2}
\end{equation}
where $\varepsilon_0$ is a constant spinor.   In particular (\ref{resvar2}) is also required by the fact that (\ref{Kvec1}) gives the time-like Killing vector.

The results of these supersymmetry variations closely parallel the computations for the $T^6$ compactification.  However, there are potentially dangerous extra terms that cancel as result of anti-self-duality of the tensors involved and the identity (\ref{asd}).

The last set of supersymmetry variations are the ones parallel to $\cM_{GH}$ and because these are quite non-trivial, we  describe them rather explicitly.   In the frame direction labelled by $c$, and using $W=Z$, we have: 
\begin{equation}
\begin{aligned}
 \coeff{1}{2}\, Z^{-1/2}\, {e^\mu}_\alpha \, \bigg[ & S_{I\, ac} \, \partial_\mu  u^I \, \Gamma^{a+4 \, \alpha} ~+~    \coeff{1}{2}\, (\partial_\mu  \log (Z/Z_0)) \,\Gamma^{c+4 \, \alpha} ~-~    \coeff{1}{6}\,  (\partial_\mu  \log Z_3 )\, \Gamma^{c+4 \, 0  \alpha\, 9\, 10} \\ 
&~-  \coeff{1}{6}\, \big (\coeff{1}{2}(\partial_\mu  \log Z_0 ) \,J_{ab} -  \, J_{a d} {S_{I}}{{}^{d}}_b  \,\partial_\mu  u^I \, \big)\, \Gamma^{c+4 \, 0  \alpha\, a+4 \, b+4 }  \\
&~+  \coeff{1}{3}\, \big ((\partial_\mu  \log Z_0) \, {J^c}_a ~-~2\,{J^c}_d \,S_{I}{{}^{d}}_a \,\partial_\mu  u^I \,   \big)\, \Gamma^{0  \alpha\, a+4 }  \bigg]\, \varepsilon\\
&\qquad \qquad\qquad\qquad\qquad~+~    \coeff{1}{4}\, Z^{-1/2}\, Z_0^{1/2}\,\, \medtilde \omega_{c \, a b} \,\Gamma^{a+4 \, b+4 } \, \varepsilon~=~ 0 \,.
 \label{vareqn4}
 \end{aligned}
\end{equation}
The last term vanishes because of (\ref{asd}) and the anti-self-duality of $ \medtilde \omega_{c \, a b}$ in the indices $a,b$. It is tempting to try to eliminate the fifth term in the same manner because of the  the anti-self-duality of  $\frac{1}{2} \omega_{I\, ab}= J_{a d} {S_{I}}{{}^{d}}_b$.  This is incorrect because of the skew-symmetrization in all three indices $[abc]$ on the $\Gamma$-matrices.  To handle the middle line in  (\ref{vareqn4}), one must first use the second projection condition in (\ref{susyproj2}) to write:
\begin{equation}
\Gamma^{c+4 \, a+4 \, b+4 }\, \varepsilon ~=~ \epsilon^{abcd} \, \Gamma^{d+4} \, \varepsilon \,,
 \label{susyproj3}
\end{equation}
The expression now involves $\Gamma^{ 0  \alpha\, d+4  }$, and to simplify this one uses the  projection conditions in (\ref{susyproj2}), to obtain: 
\begin{equation}
\Gamma^{0 \, d+4 }\, \varepsilon ~=~ -{J^d}_b \, \Gamma^{b+4} \, \varepsilon \,,
 \label{susyproj4}
\end{equation}
Finally, one uses either the self-duality of $J$, or the anti-self-duality of $\omega_I$ to get rid of the $\epsilon^{abcd}$.  In this way, one  arrives at the identity:
\begin{equation}
\begin{aligned}
 {e^\mu}_\alpha \,& \big (\coeff{1}{2}\,(\partial_\mu  \log Z_0 ) \,J_{ab} -   \, J_{a d} {S_{I}}{{}^{d}}_b  \,\partial_\mu  u^I \, \big)\, \Gamma^{c+4 \, 0  \alpha\, a+4 \, b+4 } \, \varepsilon   \\
 & ~=~ {e^\mu}_\alpha\, \big[ (\partial_\mu  \log Z_0 ) \, \Gamma^{c+4 \,  \alpha}  ~-~ 2\,\partial_\mu  u^I \, J^{ce}\, J_{db} \, {S_{I}}{{}^{d}}_e \, \Gamma^{b+4 \,  \alpha} \big] \, \varepsilon\\
 &  ~=~ {e^\mu}_\alpha\, \big[ (\partial_\mu  \log Z_0 ) \, \Gamma^{c+4 \,  \alpha} ~+~ 2 \,\partial_\mu  u^I \, {S_{I}}{{}^{c}}_b \, \Gamma^{b+4 \,  \alpha} \big] \, \varepsilon\,,
 \end{aligned}
 \label{vareqn5}
\end{equation}
where the last equality holds because of the skew-symmetry of $\frac{1}{2} \omega_{I\, be}= J_{bd} \, {S_{I}}{{}^{d}}_e  = - J_{ed} \, {S_{I}}{{}^{d}}_b$.

Finally, one uses  (\ref{susyproj4}) to simplify the last terms in  (\ref{vareqn4}) and collects everything together.  The terms involving ${S_{I}}{{}^{c}}_b \, \Gamma^{b+4 \,  \alpha}  \, \varepsilon$ cancel directly with one another, and the other terms are simply:
\begin{equation}
\Big( \coeff{1}{12}\, Z^{-1/2}\,  {e^\mu}_\alpha \,\Gamma^{c+4 \,  \alpha} \, \varepsilon \Big) \,\big (3 \, \partial_\mu  \log Z  - 2\, \partial_\mu  \log Z_0- \partial_\mu  \log Z_3 \big)\, 
 \label{vareqn6}
\end{equation}
which vanishes by virtue of  (\ref{resvar1}).

There are some important messages coming from this detailed analysis.  First, the cancellation of the various terms in the BPS equations makes heavy use of the relationship, (\ref{KahlerDefs}) or (\ref{KahlerDefsGH}), between the Lichnerowicz modes and the harmonic forms.  The former arise in the gravitino variation via the spin connections while the latter arise in the Maxwell flux, (\ref{F4flux}).  The non-trivial fibration form of the metric, (\ref{elevenmetfibered}),  using the compensators, is essential to the correct, and non-singular, appearance of the Lichnerowicz modes  and harmonic forms in the gravitino variation and is thus  essential to satisfying the BPS equations.  Finally, we have arrived at the primary result of this paper: the BPS equations are identically satisfied by our Ansatz and there is nothing perturbative about this result: it is exact.

\section{Equations of motion}
\label{sect:EOMs}

\subsection{The BPS system, equations of motion and M2-brane sources}
\label{ss:BPS-EOMs}

Since the commutator of two supersymmetries generates the Hamiltonian of a system, it follows that the integrability condition of the BPS system should lead to at least some of the equations of motion.  In particular, solving the BPS equations means that one has automatically solved at least a subset of the equations of motion. In some contexts, solving the BPS equations actually leads to solving all the equations of motion but this is not true in general.   For eleven-dimensional supergravity this was investigated in some detail in \cite{Gauntlett:2002fz}, where it was shown that the BPS integrability conditions are:
\begin{equation}
\begin{aligned}
E_{MN}\,\Gamma^{N}\,\varepsilon & ~-~ \coeff{1}{6\cdot 3!}\,*(d*F~+~ \coeff{1}{2}\, F \wedge \, F)_{P_1P_2 P_3}(\Gamma_{M}{^{P_1P_2 P_3}}-6\delta^{P_1}_{M}\Gamma^{ P_2 P_3})\,\varepsilon \\ 
& ~-~ \coeff{1}{6!} \, dF_{P_1 P_2 P_3 P_4 P_5}(\Gamma_{M}{^{P_1 P_2 P_3 P_4 P_5}} -10 \, \delta_{M}^{P_1}\Gamma^{P_2 P_3 P_4 C_5})\,\varepsilon ~=~0 \,,
\end{aligned}
\label{intcond}
\end{equation}
where 
\begin{equation}
E_{MN} ~\equiv~ \big[ R_{MN}-\coeff{1}{12}(F_{M P_1 P_2 P_3} \, {F_{N}}{^{P_1P_2 P_3}}~-~ \coeff{1}{12}\, g_{MN}\, F^2) \big ] \,.
\label{EMNdefn}
\end{equation}
It follows that, if $F$ satisfies its Bianchi identity and its equation of motion, then one must have:
\begin{equation}
E_{MN}\,\Gamma^{N}\,\varepsilon ~=~0 \,.
\label{EMNvan}
\end{equation}
By making different contractions with (\ref{EMNvan}), it is then shown in \cite{Gauntlett:2002fz} that if $K^M$, defined in (\ref{Kvec1}), is timelike, then $E_{MN} =0$ for all $M,N$.  In other words, the BPS equations, the Maxwell equations and the Bianchi identities necessarily imply the Einstein equations.  

As we have shown in Section \ref{sect:Comp}, our projection conditions imply that $K^M$ is  time-like and so, having solved the BPS equations, it suffices to check the Maxwell equations and Bianchi identities.  Given our Ansatz, (\ref{C3Ansatz}), the latter are automatically satisfied, and so it remains to examine the Maxwell equations, which we will do in this section. 

One should note that the proof in  \cite{Gauntlett:2002fz} was done purely for the eleven-dimensional supergravity action, without explicit brane sources.  We are  going to assume the corresponding result {\it with} M2-brane sources. While this has not been explicitly proven, what matters is how the BPS integrability condition incorporates the inclusion of brane sources in a supersymmetric action.  The structure of those integrability conditions must simply add extra source terms to each of the terms in the first line of (\ref{intcond}).  Thus, even with supersymmetric M2-brane sources, solving the BPS equations and satisfying the Bianchi identities for $F$, will imply that solving the Maxwell equations with M2-brane sources will mean that the Einstein equations, with the corresponding sources, will also be satisfied.

Another issue that will arise is the smearing of the M2-branes within the full, eleven-dimensional solution. In a Calabi-Yau compactification, wrapping M2-branes on $2$-cycles in $\cY$, gives rise to electric-charge sources in the effective five-dimensional theory.  The electric field lines of these  sources can only extend to infinity in the five-dimensional space-time, $\cX$, and so, on scales much larger than the compactification scale,  the electric potentials fall off as $r^{-2}$ in the space-time. If one probes the solution on scales less than the compactification scale, one should expect to see deviations from the effective field theory and see details of the M2-brane distribution in the full eleven-dimensional geometry.   

For  torus compactifications, the issue of effective field theories can be obviated through consistent truncation.  If the compactification manifold has a transitive symmetry group, then it is always a consistent truncation if one restricts to all the fields that are independent of the compactification manifold.  That is, the reduced theory in lower dimensions is not merely an effective theory, it is actually a consistent truncation in that solving the lower-dimensional  equations of motion yields an {\it exact} (as opposed to approximate or effective) solution to the higher-dimensional equations of motion.  To incorporate branes in such a consistent truncation, they must be uniformly smeared over the compactification directions that are transverse to the branes.  This will preserve the transitive symmetry on all the compactification directions and thus incorporate such smeared brane sources within the consistent truncation.  For the M2-branes wrapping a $T^2$ inside a $T^6$ compactification, the branes can be uniformly smeared over the transverse $T^4$. In a flat spatial base, $\cB$, this leads to a pure $r^{-2}$ behavior and the brane sources can be concentrated into a delta-function on $\cB$.   For non-trivial compactifications on manifolds without transitive symmetries, like $\cM_{GH} \times T^2$,  one must necessarily return to either using an effective field theory that is valid on scales much larger that the compactification scale, or, if one seeks a solution to the eleven-dimensional system, one must make choices of brane distributions on the compactification manifold and solve the eleven-dimensional equations with those brane sources.  

We now investigate these issues in a little more detail and, for simplicity, we will take $\cB$ to be flat $\IR^4$ in the rest of this paper.

\subsection{The Maxwell equations}
\label{ss:MaxEqns}

Since we are using an electric Ansatz for the fluxes, (\ref{C3Ansatz}),  one has $F \wedge F = 0$  and the left-hand side of the  Maxwell equations becomes:
\begin{equation}
	\begin{split}
	d \ast F^{(4)} ~=~ \mbox{vol}_4 \wedge \big[ \Box{Z_0}\, J & + (2 \partial_{\mu}Z_0 \,\partial^{\mu} u^I  + Z_0\, \Box u^I ) \, \omega_I + Z_0\, \partial_{\mu} u^I\partial^{\mu} u^J\, \cD_J \omega_I  \big] \wedge \mbox{vol}_{T^2} \\
	& -\left(\frac{Z_0}{Z_3}\right)^{2/3} \Box{Z_3}\, \mbox{vol}_4 \wedge \mbox{vol}_4^{'} \,,
		\end{split}
		\label{dstarF}
\end{equation}
where $\Box$ is a standard Laplacian on $\cB = \mathbb{R}^4$,  and $\mbox{vol}_4$ and $ \mbox{vol}_4^{'}$ are the volume forms on $\cB$ and $\cM_{GH}$.   
Note also that, because of the compensators, the exterior derivative on $\omega_I$ appearing in $F^{(4)}$ (see (\ref{F4flux})) has been promoted to  $\cD_J \omega_I$.  One can easily evaluate this explicitly and the results depends on the details of the harmonic forms:
\begin{equation}
	\cD_J \omega_I = \bigg[ \partial_a\bigg( \frac{1}{V} \partial_3 \left( \frac{K^I K^J}{V} -\delta_{I J} K^I \right)\bigg)\bigg] \Omega^{(a)}_{-} + \bigg(\partial_3 \left( \frac{K^J}{V} \right)\bigg)\, \omega_I + \vec{\nabla} \left( \frac{K^I}{V} \right)\cdot \vec{\nabla} \left( \frac{K^J}{V} \right) (J-\Omega^{(3)}_{-}) \,.
\end{equation}

The electrostatic potential, $Z_3$, behaves exactly as it does in the $T^6$ compactification: It is a harmonic function and decouples from other scalars.  We will therefore make the standard, spherically symmetric choice:
\begin{equation}
	Z_3 ~=~ 1+\frac{Q_3}{r^2}\,.
\end{equation}
The interesting new features are associated with $\cM_{GH}$ and its modulus and so we focus on these.  

First, for small fluctuations of $u^I$ about the their stationary values, we can neglect the terms that are quadratic in $\partial_\mu u^I$ in the Maxwell equations. Since  $J$ and the $\omega_I$ are  linearly independent and provide a basis of harmonic $(1,1)$-forms,   one obtains a set $N$  equations (to linear order in fluctuations):
\begin{equation}
	\Box{Z_0} ~=~ 0\,,  \qquad 2 \partial_{\mu}Z_0 \,\partial^{\mu} u^I  + Z_0\, \Box u^I ~=~ 0\,, \quad I=1\ldots N-1\,,
	\label{linsol1}
\end{equation}
or
\begin{equation}
	\Box{Z_0} ~=~ 0\,,  \qquad \Box{(Z_0\, u^I) } ~=~ 0\qquad \Rightarrow  \qquad u^I ~=~  \frac{Z_I}{Z_0} \,, \quad \Box{Z_I} ~=~ 0  \,.
	\label{linsol2}
\end{equation}
Thus the moduli are ratios of harmonic functions.  If we further impose spherical symmetry  one obtains\footnote{One can scale the harmonic functions by a constant and in this way one can give the $Z_I$ and $u^I$ an overall scale.  We have chosen to make the moduli, $u^I$, dimensionless.}: 
\begin{equation}
	Z_0 ~=~ 1+\frac{Q_0}{r^2}\,, \qquad Z_0 \,u^I~=~ 1+\frac{Q_I}{r^2} \qquad \Rightarrow  \qquad u^I ~=~ \frac{r^2+Q_I }{r^2+Q_0} \,.
	\label{linsol3}
\end{equation}
As one would expect, these results correspond precisely with the effective five-dimensional field theory that emerges from a compactification when the manifold, $\cY$,  is small.  Indeed, to arrive that this result we merely assumed that  the fluctuations in the moduli, $u^I$, were small.  Note also that the moduli undergo the expected attractor behavior between $r=0$ and $r= \infty$.

Moving away from slowly varying moduli and effective field theory, we can consider  (\ref{dstarF}) in general and see what it implies for eleven-dimensional BPS solutions.   

One can define a three-form charge density, $\lambda$, as the right-hand side of the Maxwell equations:
\begin{equation}
	\ast d \ast F^{(4)} ~\equiv~ - \lambda\,.
	\label{M2density}
\end{equation}
In principle, one can then allow any choice for the fields $Z_0$ and $u_I$ and this will lead to a BPS solution to the equations of motion  for some distribution of M2 branes.  In practice, we want to find BPS solutions for rather more physical choices of brane distributions. The obvious choice is to consider the solution to some form of five-dimensional, effective field theory and determine the corresponding distributions of M2 branes in  eleven-dimensional supergravity.   The best, most canonical choice is to render $\ast \lambda$ topologically trivial on $\cM_{GH}$.  That is, we take the right-hand side of (\ref{dstarF}) and project onto all the harmonic  forms\footnote{Here we mean harmonic on  $\cM_{GH}$, neglecting any dependence of the moduli on the space-time.}, which will include extracting the harmonic pieces, proportional to $J$ and each $\omega_I$,  from  $\cD_J \omega_I$.   If one sets the coefficients of the harmonic projections of (\ref{dstarF})  to zero, the result will be $N$ equations for $Z_0$ and the $u^I$ and, unlike (\ref{linsol1}), these equations will include the $\partial_{\mu} u^I\partial^{\mu} u^J$ terms.  We will refer to this as the {\it topological solution}.    Note that, depending on the harmonic content of $\cD_J \omega_I$, it may differ from the solution to (\ref{linsol1}) at the non-linear level.

Solving such a homogeneous system really means that one is choosing a $\delta$-function source  for the brane charges at some location in $\cB$.  In (\ref{linsol3}), the branes are all located in a $\delta$-function source  at $r=0$.  Away from such as $\delta$-function source, the topological solution means that $*\lambda$ is {\it exact}.  As a result, any integral of  $\ast F^{(4)}$ over a Gaussian surface that excludes the $\delta$-function source will be zero and thus the M2-brane charge density represented by  $*\lambda$ is entirely dipolar outside the  $\delta$-function source.   In addition,  the projection of  $*\lambda$ on $\cM_{GH}$ is also exact.  Consequently, for any point in $\cB$ that lies outside  $\delta$-function source, the charge distribution on the $\cM_{GH}$ fibered above that point has no net charge on any cycle in $\cM_{GH}$ and so, once again, the charge distribution is dipolar. 

For torus compactifications, or any other compactification manifold with a transitive isometry group, the M2-brane charge density, $\lambda$, can be smeared so that the dipolar distribution is replaced by its average, and hence is exactly zero outside the $\delta$-function source.  The cost of using more general compactification manifolds is that the brane distribution is typically non-zero but {\it entirely dipolar} on the compactification fibers outside the $\delta$-function source.  This means that if one looks on scales much larger than the compactification, the dipolar distributions average to zero and the non-zero brane charges are only localized at the $\delta$-function source in $\cB$, as required by effective field theory. Zooming in on the compactification scale, one sees how the supersymmetric topological solution resolves into dipolar brane distributions outside the source.

More generally, given that the M2-brane charge densities do not necessarily vanish  outside the $\delta$-function sources in $\cB$, one can also consider other supersymmetric, physical brane distributions that will reduce to something similar to the results one gets from the topological solution.  For example, rather than concentrating the brane charge entirely in a $\delta$-function  in $\cB$, one can spread the brane charge out in $\cB$ on the scales comparable to that of the compactification and, in so doing, rearrange the dipole densities in $\cM_{GH}$.  If one wants to go beyond the topological solution, there are many choices and we will not dwell upon them here.  We will examine this further by computing an explicit example.

\section{An example: Eguchi-Hanson}
\label{sect:EH}

We now consider compactifying on the two-centered Gibbons-Hawking metric with unit charges,  commonly known as the Eguchi-Hanson (EH) metric.  This metric can be obtained as a hyper-K\"ahler resolution of the conical singularity of $\IC^2/Z_2$ and depends on the blow-up parameter $a$, which is the only modulus.  One can also think of the EH metric as a ``local model''   of a $\rm K3$ metric near a single, isolated $2$-cycle.  

The harmonic $(1,1)$-form, $\omega$, is dual to the blown-up cycle and, together with the complex structure, $J$, forms a basis of the cohomology.  Moreover, since $J \wedge \omega =0$, the Maxwell fields arising from this compactification have similarly trivial intersection and thus we get simple $U(1) \times U(1)$ gauge fields.  Including the Maxwell field from the $T^2$ factor, we obtain the standard ``STU''  effective field theory in five dimensions.  Indeed, the Eguchi-Hanson compactification closely parallels the role of the $T^4$ (or, more precisely, the $\IR^4$) in the $T^6$ compactification discussed in Section \ref{ss:T6comp}.  For the $T^6$ compactification we can make this parallel more precise by noting that
\begin{equation}
J  ~=~  u\, dy^1 \wedge dy^4 ~+~       u^{-1} \,  dy^2 \wedge dy^3 \,, \qquad \omega  ~=~  u\,  dy^1 \wedge dy^4 ~-~  u^{-1} \,    dy^2 \wedge dy^3  \,.
\label{JomegaT4}
\end{equation}
defines a self-dual complex structure and an anti-self-dual harmonic form on $T^4$.  Moreover, one also has $J \wedge \omega =0$ and, from the perspective of  effective field theory, the compactifications on $T^6$ and $\cM_{EH} \times T^2$ yield the same five-dimensional  theory.

\subsection{Complex coordinates and trivializing the compensators}
\label{ss:CCtriv}

The two-centered EH metric has a lot more symmetry that a general GH metric and so, rather than use the coordinates and frames described in Section \ref{ss:GHredn}, we will use complex coordinates, $(z_1,z_2)$,  adapted to the $SU(2) \times U(1)$ symmetry and K\"ahler structure of the metric:
\begin{equation}
	z_1 ~=~ \rho \cos\left(\frac{\theta}{2}\right) e^{\frac{i}{2}(\psi+\phi)}\,, \qquad z_2 ~=~ \rho \sin\left(\frac{\theta}{2}\right) e^{\frac{i}{2}(\psi-\phi)}\,.
	\label{compcoords}
\end{equation}
We will also see that this choice of coordinates greatly simplifies the compensators and related structure that we described in Section \ref{ss:GHcomps}.
 
The K\"ahler potential is given by
\begin{equation}
	K(\rho^2) ~=~ \sqrt{\rho^4+a^4} + a^2 \log\left(\frac{\rho^2}{\sqrt{\rho^4+a^4}+a^2}\right)\,,
\end{equation}
and the K\"ahler form is
\begin{equation}
	J ~=~ \frac{i}{2} \partial \bar{\partial} K(\rho^2) ~=~ \frac{\rho^3}{\sqrt{\rho^4+a^4}}\, d \rho \wedge \sigma_3 + \sqrt{\rho^4+a^4}\, \sigma_1 \wedge \sigma_2 \,,
	\label{Kform-old}
\end{equation}
where the Pauli matrices are  
\begin{equation}
	\sigma_1 = \frac{1}{2} (\sin\psi\, d\theta - \sin\theta \cos\psi\, d\phi) \,, \quad \sigma_2 = -\frac{1}{2} (\cos\psi\, d\theta + \sin\theta \sin\psi\, d \phi) \,, \quad  	\sigma_3 = \frac12(d\psi+\cos\theta\, d\phi)\,.
\end{equation}
The metric is given by
\begin{equation}
	\medtilde{ds}^2 ~=~ \frac{1}{\sqrt{1+\frac{a^4}{\rho^4}}} \left( d \rho^2 +\rho^2 \sigma_	3^2 \right) + \rho^2 \sqrt{1+\frac{a^4}{\rho^4}} \left(\sigma_1^2 + \sigma_2^2 \right) \,,
	\label{EHmet}
\end{equation}
and harmonic two-form may be written:
\begin{equation}
	\omega ~=~ -\frac{a^2 \rho^3}{(\rho^4+a^4)^{3/2}}\, d\rho \wedge \sigma_3 + \frac{a^2}{\sqrt{\rho^4+a^4}}\, \sigma_1 \wedge \sigma_2\,.
	\label{harmEH}
\end{equation}
We will find it convenient to identify the dimensionless modulus, $u$, as
\begin{equation}
	u ~\equiv~ a^2/\Lambda_h^2 \,,
	\label{modulus}
\end{equation}
where $\Lambda_h$ is a length scale on $\cM_{EH}$.  This length scale is that of the (compact) homology cycles and, for consistency, it must be very much less than any cut-off, $\Lambda_\infty$, that one uses to regulate fields at infinity in the EH manifold.

One can  easily check that
\begin{equation}
	\frac{\partial J}{\partial u} ~=~\Lambda_h^{2}\,  \frac{\partial J}{\partial a^2} ~=~\Lambda_h^{2}\,  \omega\,.
\end{equation}
It follows that the Lichnerowicz mode coming from the  K\"ahler deformation of the metric generated by $\omega$ is identically equal the derivative of metric with respect to $a^2$.  The compensator fields are therefore trivial:  $B^i \equiv 0$ and $\cD =\Lambda_h^2 \partial_{a^2}$.   As we noted at the end of Section \ref{ss:GHcomps}, the ``curvatures,'' ${Y^i}_{I J}$,  vanish identically and so one might anticipate from this  ``flatness'' that one should be able find a way to trivialize the ${B^i}_I$.  The  complex coordinates, (\ref{compcoords}), achieve this for the EH metric.  

Finally, the  derivative of $\omega$ is simply 
\begin{equation}
	\Lambda_h^{-2}\, \cD \omega ~=~ \partial_{a^2} \omega ~=~ -\frac{\rho^3 (\rho^4-2 a^4)}{(\rho^4+a^4)^{5/2}}\, d\rho \wedge \sigma_3 +\frac{\rho^4}{(\rho^4+a^4)^{3/2}}\, \sigma_1 \wedge \sigma_2 \,.
	\label{dEHharm}
\end{equation}
It is easy to verify that this has a potential: 
\begin{equation}
	 \cD \omega ~=~ d_y \eta \,, \qquad{\rm where} \qquad \eta ~\equiv~   \frac{\Lambda_h^{2}\,  \rho^4}{2 (\rho^4 + a^4)^{3/2}} \, \sigma_3   \,.
	 \label{Domexact}
\end{equation}
The $1$-form, $\eta$, is ``compact'' in that it has finite norm and vanishes at infinity.  Thus $ \cD \omega$ is indeed cohomologically trivial on the EH space.  

Note that these statements about exactness apply only within the cohomology of the  EH space:  If one replaces $d_y$ by $d$ and allows $a$ to become a function of $x$, then (\ref{Domexact}) is no longer true.

Finally, we note the major difference between the harmonic forms on $T^4$ and the EH manifold.  Observe that (\ref{JomegaT4}) implies
\begin{equation}
	u  \partial_u \, \omega  ~=~ J \,,
\label{T4diff}
\end{equation}
which should be contrasted with  (\ref{dEHharm}) and (\ref{Domexact}).  The fact that $ \partial_u \, \omega$ is cohomologically trivial in Eguchi-Hanson and non-trivial on $T^4$  leads to a very significant modification of the differential equation for $Z_0$ and thus has a significant impact on the form of the solution on the EH manifold compared to the $T^4$.

\subsection{The topological solution}
\label{ss:EHEFTsol}

Since (\ref{Domexact}) implies that $\cD \omega$ is exact, it follows that the harmonic pieces of  $d \ast F^{(4)}$ are precisely the first two terms on the right-hand side of (\ref{dstarF}).  As a result, the topological solution described in Section \ref{ss:MaxEqns} is exactly the linearized solutions given in (\ref{linsol1})--(\ref{linsol3}), except that we are no longer linearizing the result.   We therefore have:
\begin{equation}
	   \qquad a^2 ~=~ \Lambda_h^2 \, \frac{Z_1}{Z_0} \,, \qquad  \Box{Z_0} ~=~ \Box{Z_1} ~=~0  \,,
\label{effsol1}
\end{equation}
where we have set $I=1$ in (\ref{linsol3}). It follows from (\ref{Domexact}), (\ref{dstarF}) and (\ref{M2density}) that we have:
\begin{equation}
\begin{aligned}
	* \lambda  ~=~&  d\, \Big[ Z_0\, \big(\partial_{\mu} u \, \partial^{\mu}u \big)\, {\rm vol}_4 \wedge\, \eta \, \wedge {\rm vol}_{T^2} \Big]  \\
	~=~  & d\, \bigg[  Z_0\, \big(\partial_{\mu} u \,\partial^{\mu} u\big)\, \frac{\Lambda_h^{2}\,  \rho^4}{2 (\rho^4 + a^4)^{3/2}} \,  {\rm vol}_4 \wedge\,  \sigma_3 \, \wedge {\rm vol}_{T^2}  \bigg] \,.
\end{aligned}
	 \label{starlam1}
\end{equation}
For the spherically symmetric solutions in (\ref{linsol3}) this reduces to:
\begin{equation}
	Z_0 ~=~ 1+\frac{Q_0}{r^2}\,,    \qquad a^2 ~=~ \Lambda_h^2 \, u  ~=~ \Lambda_h^2 \,  \frac{r^2+Q_1 }{r^2+Q_0} \,,
\label{effsol2}
\end{equation}
with
\begin{equation}
	* \lambda ~=~   d\, \bigg[   \frac{2\,(Q_1 -Q_0)^2}{(Q_0 + r^2)^3} \, \frac{\Lambda_h^{2}\,  \rho^4}{ (\rho^4 + a^4)^{3/2}} \,  {\rm vol}_4 \wedge\,  \sigma_3 \, \wedge {\rm vol}_{T^2} \bigg] \,.
	 \label{starlam2}
\end{equation}

There are several things to note about the brane distribution encoded in $* \lambda$.  First, $*\lambda$ is exact, as it must be, and is thus intrinsically a multipolar distribution.  Indeed, the potential for $*\lambda$ falls off as $r^{-6}$ in $\cB$, as befits a higher multipole distribution of charge. In the internal EH manifold, this potential vanishes at $\rho = 0$, peaks at $\rho = 2^{1/4} a$ and falls off as  $\rho^{-2}$ at infinity and so, in the EH manifold, the brane distribution is localized around the non-trivial cycle at a scale set by $a$. 

Thus the EH manifold, unlike  $T^4$, has a non-vanishing $*\lambda$.  However, the charge distribution that it represents involves only higher multipoles that localize in a restricted region of $\cM_{GH}$ and very close to $r=0$ in $\cB$.

\subsection{Another effective action}
\label{ss:EHeffectiveaction}

The bosonic action for the ``STU'' model has the form:
\begin{eqnarray}
  S = \frac {1}{ 2 \kappa_{5}^2} \int\!\sqrt{-g}\,d^5x \Big( R  -\coeff{1}{2} Q_{IJ} F_{\mu \nu}^I   F^{J \mu \nu} - Q_{IJ} \partial_\mu X^I  \partial^\mu X^J -\coeff {1}{24} C_{IJK} F^I_{ \mu \nu} F^J_{\rho\sigma} A^K_{\lambda} \bar\epsilon^{\mu\nu\rho\sigma\lambda}\Big) \,,
  \label{5daction}
\end{eqnarray}
where $I, J =1,2,3$ and $C_{IJK} = | \epsilon_{IJK} |$. The  scalars satisfy the constraint $X^1 X^2 X^3 =1$ and metric for the kinetic terms is:
\begin{equation}
 Q_{IJ} ~=~ \frac{1}{2}\,{\rm diag}\,\left((X^1)^{-2} , (X^2)^{-2},(X^3)^{-2} \right) \,.
\label{scalarkinterm}
\end{equation}
The standard way of parametrizing the scalars and the warp factor in the five-dimensional metric is to introduce three scalar fields, $Z_I$, and write:
\begin{equation}
	Z  ~\equiv~  (Z_1 Z_2 Z_3)^{1/3} \,, \qquad X^J   ~\equiv~ \frac{Z}{Z_J}	 \,.
\label{XZrelns}
\end{equation}

We can  arrive at another form of the effective action by simply focussing on the scalar kinetic terms.  We first eliminate $X^3$ using  $X^3 = (X^1 X^2)^{-1}$ and define the independent dynamical degrees of freedom:   $A=(X^1 X^2)^{1/2}$ and $B =(X^2/X^1)^{1/2}$.  On the $T^4$, the scalar, $A$,  measures its volume relative to the $T^2$, while $B$ determines the relative size of  $T^2 \times T^2$ factors of the $T^4$.   Comparing with the compactification on $\cM_{GH} \times T^2$,  $A$ is the analog of $Z_0$ while $B$ is the analog of $a^2$.   Expressing the scalar kinetic term in (\ref{5daction}) using the fields $A$ and $B$, we find: 
\begin{equation}
	S_X ~=~\int\!\sqrt{-g}\,d^5x \left(  \frac{3\,(\partial_{\mu} A)^2}{A^2} + \frac{(\partial_{\mu} B)^2}{B^2} \right) \,.
\end{equation}

On the other hand, one can start with the Einstein action in eleven-dimensional supergravity and compute the scalar kinetic terms using the metric Ansatz (\ref{11dmetric}).  To do this, one must integrate over the non-compact $\cM_{EH}$ and this requires a cut-off.  We simply choose to integrate $\rho$ from $0$ to the scale, $\Lambda_\infty$.  To simplify this computation we note that we can obtain the form of the scalar kinetic terms by restricting the scalar to be a function of one variable on $\cB$.  We take this variable to be the radial coordinate, $r$.  The result of this computation is:
\begin{equation}
	S_{scalar} ~=~\int\!\sqrt{-g}\,d^5x \left(  \frac{3\, (\partial_r X^0)^2}{(X^0)^2} + \frac{4 a^2 (\partial_r a)^2}{\Lambda_\infty^4 + a^4} \right) \,,
\end{equation}
where $X^0=Z/Z_0$. 

This suggests the following identification between moduli of the our model and the moduli of the STU model\footnote{Obviously, given that we are matching quadratic forms, one can make an arbitrary rotation of the identification we are using.}:
 \begin{equation}
		X^0 ~\leftrightarrow~ A~=~\sqrt{X^1 X^2}\,, \qquad \int\frac{2 a \partial_r a}{\sqrt{\Lambda_\infty^4+a^4}}\, dr ~=~ \log(a^2+\sqrt{\Lambda_\infty^4+a^4}) ~\leftrightarrow~ \log B ~=~\frac12\log \frac{X^2}{X^1} \,. 
 \end{equation}
where we have fixed the constant of integration by taking $B=1$ when $X^2 = X^1$. One obtains the map
\begin{equation}
	Z_0 ~=~ \sqrt{Z_1 Z_2} \,, \qquad  a^2 ~=~ \frac12 \Lambda_\infty^2 (B-B^{-1}) ~=~  \frac12 \Lambda_\infty^2 \frac{Z_1^2 - Z_2^2}{Z_1 Z_2} \,,
	\label{Zamap}
\end{equation}
for $Z_1^2  \ge Z_2^2$.

One can make a similar reduction of the eleven-dimensional Maxwell equations to get the five-dimensional effective action of these degrees of freedom. To do this we evaluated the integral of $\ast F^{(4)} \wedge F^{(4)}$ in eleven dimensions for our flux Ansatz (\ref{F4flux}) and we find:
\begin{equation}
	S_{Maxwell} ~=~\int\!\sqrt{-g}\,d^5x \left( \frac{(\partial_r Z_0)^2}{(Z_0)^2} + \frac{4 a^2 (\partial_r a)^2}{\Lambda_\infty^4 + a^4} + 			\frac12 \frac{(\partial_r Z_3)^2}{(Z_3)^2}  \right) \,. 
\end{equation}
With the help of map (\ref{Zamap}) this can be rewritten as 
\begin{equation}
	S_{Maxwell} ~=~\int\!\sqrt{-g}\,d^5x \left( \frac12 \sum_{I=1}^{3} \frac{(\partial_r Z_I)^2}{(Z_I)^2}\right) \,.
\end{equation}
This is exactly what one will get from the action of STU model (\ref{5daction}) if one substitutes the electric Ansatz for five-dimensional Maxwell fields
\begin{equation}
	F_I~=~ - d ( Z_I^{-1} \, dt) \,.
\end{equation}

The five-dimensional theory implies that the functions, $Z_I$, are harmonic, and for spherically symmetric solutions they can be chosen as in (\ref{linsol3}).  The identification (\ref{Zamap}) leads to a different eleven-dimensional solution to the one given in Section \ref{ss:EHEFTsol}

While this attempt at arriving at an effective action looks a little more like the $T^4$ effective action, particularly with the form of $Z_0$ in   (\ref{Zamap}), it is of more limited utility than the solution in Section \ref{ss:EHEFTsol}.  This is because we already have requirement $Z_1 \ge Z_2$  and we must also keep $ a \ll \Lambda_\infty$ for consistency.   This means we must have requires  $B \sim 1$, or $\frac{Z_1}{Z_2} \approx 1$. 
Similarly, the identification (\ref{Zamap}) means that the coefficients of $J$ and $\omega$ in  $\ast \lambda$ do not vanish and so there are non-trivial monopolar charge sources smeared on $\cB$.  This smearing in $\cB$ will only remain small if $\frac{Z_1}{Z_2} \approx 1$.  With such restrictions we are  essentially back in the linearized regime in which the fluctuations of $u$ must remain small. 

Here we have taken an obvious but rather `ad hoc' procedure to arrive at an effective action and a solution.  The issue with such a generic approach is that it will typically have a limited range application and lead to non-trivial smearing of charge sources in the space-time.  In contrast, the topological solution described in Sections \ref{ss:MaxEqns} and  \ref{ss:EHEFTsol} produces solutions whose charge distribution is intrinsically dipolar on the $\cY$ fibers outside the desired space-time sources and that remains valid for generic solutions of the underlying equations of motion in  the effective field theory. 

\section{Conclusions}
\label{sect:Conc}

String theory makes extensive use of non-trivial Calabi-Yau compactifications to reduce M theory, or type II supergravity, to five, or four, dimensions.  Usually such compactifications are used to obtain an effective field theory  in the lower dimension and the back-reaction of the fluxes on,  and deformations of, the Calabi-Yau are typically ignored.  However, as we attempt a deeper and more sophisticated understanding of string theory compactifications, understanding these back-reactions and the dynamical effects of exciting fluxes and moduli become ever more important.  There are several research programs that are driving this need, including holographic field theory, string cosmology and black-hole microstate structure.  Our particular interest comes from the latter and particularly upon the description of supersymmetric microstate structure within the microstate geometry program. 

For much of its evolution, the study of microstate geometries has focussed on geometric transitions in the space-time directions and the topology and geometric dynamics of the compactification manifold has been secondary.  Indeed, most of the exact supergravity computations have involved trivial, toroidal compactifications.  As we described in the Introduction, there are a growing number of reasons to investigate the dynamics of the internal degrees of freedom at the same level of detail as we have investigated the space-time dynamics and, more broadly, put the space-time manifold and the compactification manifold on the same footing.  

This means that we must replace the toroidal compactification with non-trivial Calabi-Yau manifolds whose homology cycles can blow up or blow down without the volume becoming singular.  In this paper we have used a Gibbons-Hawking ALE space, $\cM_{GH}$, as a local model of a K3 surface and we have found a complete solution to the eleven-dimensional supersymmetry conditions (BPS equations) for a compactification on $\cM_{GH} \times T^2$ that involves purely electric charges.  We also required that the supersymmetries have exactly the same structure (with the same projectors) as the three-charge black hole in five dimensions.  This solution to the BPS equations requires that we fiber the compactification manifold over the space-time base using the compensator fields on $\cM_{GH}$ and then, having done that, the solution to the BPS equations is relatively straightforward.

Solving the BPS equations will generically only solve a subset of the equations of motion. We argued, following, \cite{Gauntlett:2002fz}, that solving the Bianchi identities and the Maxwell equations in addition to the BPS equations, even in the presence of brane sources, implies the Einstein equations are satisfied.  We therefore examined the Maxwell equations.  This also highlights a physical issue with non-trivial Calabi-Yau compactifications: In toroidal compactifications one smears the brane distributions over the compactification directions that lie transverse to the branes whereas a generic Calabi-Yau manifold does not possess a set of transitive symmetries transverse to the branes and so there is no canonical definitions of a ``uniform'' distribution of branes in such transverse directions.  Instead one must examine choices of solutions and the brane distributions to which they lead.  The most canonical is the ``topological choice,'' in which the solutions are chosen so as to eliminate all the sources proportional to the harmonic forms on the internal manifold.  This is the most natural choice for several reasons.  First, this is the choice motivated by the effective five-dimensional field theory that emerges from the topology of the internal manifold.  Secondly,  making the charge density exact on the internal manifold means that it carries no net charge and is therefore intrinsically dipolar (outside any explicit sources in the space-time). On scales much larger than the compactification scale, the fields sourced by such a distribution fall off very rapidly.  Indeed, for the  $\cM_{EH} \times T^2$ solutions we found that this dipolar distribution  localized around the non-trivial cycle.  

There are, of course, many other choices for a supersymmetric brane distributions and some of them emerge from different perturbative expansions or other ways of arriving at a five-dimensional effective field theory, as in Section \ref{ss:EHeffectiveaction}.  However, the best way to match the physics that one expects in five, or four, dimensions is to concentrate the charge into a specified region (like a $\delta$-function support) in the lower-dimensional solution and ensure that, if there is non-zero charge density elsewhere in space then it must be intrinsically dipolar on the internal space.  In the absence of transitive symmetries along which one can smear distributions, the topological solution represents the simplest and best charge distribution for achieving this goal.

This paper obviously represents a first step in solving the larger problem of finding more general, exact compactification solutions on (local models of) non-trivial Calabi-Yau compactifications.   Here we focussed simply on electric charge sources and showed exactly how the compensators can be used to produce exact BPS solutions.  The next obvious step is to generalize this to incorporate magnetic fluxes and then try to find completely smooth exact solutions that are sourced entirely by non-trivial magnetic fluxes threading both the space-time and internal homology.  It would also be very interesting to see if one could replace our two hyper-K\"ahler four-folds, $\cB$ and  $\cM_{GH}$, in a more unified description based on a single hyper-K\"ahler eight-fold.  We would also like to see if there are new classes of solutions in which the internal compactification manifold, and even the hyper-K\"ahler eight-fold, could be ambi-polar \cite{Bena:2007kg}.  The ultimate goal of such an enterprise is to put the internal and space-time topological structure of microstate geometries on a more symmetric footing and give a more unified description of the dynamics of black-hole microstructure.

\section*{Acknowledgments}

\vspace{-2mm}
We thank Iosif Bena for valuable discussions.  We are also very grateful to the IPhT, CEA-Saclay for hospitality during  part of this project.  This work  was supported in part by the DOE grant DE-SC0011687.

\newpage
%
\begin{appendix}

\section{Supergravity Conventions}
\label{app:conventions}

The metric is  ``mostly plus,'' and we take the gamma-matrices to be
\begin{eqnarray}
\Gamma_0 &~=~&  -i \, \Sigma_2 \otimes 
\gamma_9 \,, \quad
\Gamma_1~=~  \Sigma_1 \otimes \gamma_9 \,, \quad 
\Gamma_2~=~   \Sigma_3 \otimes \gamma_9 \,,  \\
\Gamma_{j+2} &~=~&  {\oneone}_{2 \times 2} \otimes \gamma_j \,, 
\quad j=1,\dots,8 \,, 
\end{eqnarray}
where the $\Sigma_a$ are the Pauli spin matrices, $\oneone_{2 \times 2} $ is
the identity matrix, and  the $\gamma_j$ are real, symmetric $SO(8)$
gamma matrices.  As a result, the $\Gamma_j$ are all real, with 
$\Gamma_1$ skew-symmetric and $\Gamma_j$ symmetric for $j>2$.  One also
has:
\begin{equation}
\Gamma^{0 1 \cdots\cdots 10} ~=~ \oneone \,,
\label{Gammaprod}
\end{equation}
where $\oneone$  denotes the $32 \times 32$ identity matrix.

The gravitino variation is 
\begin{equation}
\delta \psi_\mu ~\equiv~ \nabla_\mu \, \varepsilon ~+~ \coeff{1}{288}\,
\Big({\Gamma_\mu}^{\nu \rho \lambda \sigma} ~-~ 8\, \delta_\mu^\nu  \, 
\Gamma^{\rho \lambda \sigma} \Big)\, F_{\nu \rho \lambda \sigma}
\label{gravvarapp}  \,.
\end{equation}
With these conventions, sign choices and normalizations,
the equations of motion are:
\begin{eqnarray}
R_{\mu \nu} ~+~ R \, g_{\mu \nu}  &~=~&  \coeff{1}{12}\, 
F_{\mu \rho \lambda \sigma}\, F_\nu{}^{ \rho \lambda \sigma}\,,
\label{EinEqn}\\ 
\nabla_\mu F^{\mu \nu \rho  \sigma} &~=~& -  \coeff{1}{1152} \, 
\varepsilon^{\nu \rho  \sigma \lambda_1\lambda_2
\lambda_3 \lambda_4 \tau_1\tau_2  \tau_3 \tau_4} \,F_{ \lambda_1  \lambda_2
 \lambda_3  \lambda_4} \, F_{ \tau_1\tau_2 \tau_3 \tau_4} \,.
\label{MaxEqnA}
\end{eqnarray}
The Maxwell equation may be written more compactly as:
\begin{equation}
d*F ~+~ \coeff{1}{2} \, F \wedge F ~=~ 0 \,. \label{MaxEqnB}
\end{equation}
%

\end{appendix}


\end{document}